\documentclass[prb,twocolumn,showpacs]{revtex4}

\usepackage{graphicx}
\usepackage{amsmath}
\usepackage{amssymb} 

\renewcommand{\vec}{\boldsymbol}

\begin{document}

\title{Electron-phonon heat transfer in monolayer and bilayer graphene}

\author{J. K. Viljas}
\affiliation{Low Temperature Laboratory, Aalto University, P.O. Box 15100, FI-00076 AALTO, Finland}
\author{T. T. Heikkil\"a}
\affiliation{Low Temperature Laboratory, Aalto University, P.O. Box 15100, FI-00076 AALTO, Finland}

\date{\today}

\begin{abstract}
 We calculate the heat transfer between electrons to acoustic and
 optical phonons in monolayer and bilayer graphene (MLG and BLG)
 within the quasiequilibrium approximation.  For acoustic phonons, we
 show how the temperature-power laws of the electron-phonon heat
 current for BLG differ from those previously derived for MLG and note
 that the high-temperature (neutral-regime) power laws for MLG and BLG
 are also different, with a weaker dependence on the electronic
 temperature in the latter.  In the general case we evaluate the heat
 current numerically.  We suggest that a measurement of the heat
 current could be used for an experimental determination of the
 electron-acoustic phonon coupling constants, which are not accurately
 known.  However, in a typical experiment heat dissipation by
 electrons at very low temperatures is dominated by diffusion, and we
 estimate the crossover temperature at which acoustic-phonon coupling
 takes over in a sample with Joule heating. At even higher
 temperatures optical phonons begin to dominate.  We study some
 examples of potentially relevant types of optical modes, including in
 particular the intrinsic in-plane modes, and additionally 
 the remote surface phonons of a possible dielectric substrate.
\end{abstract}


\maketitle


\section{Introduction}

In the presence of Joule heating, the static temperature of the
conduction electrons of a metallic system is set by the heat balance
between the heating and energy relaxation. The latter is caused either
by electron diffusion away from the heated region or by the energy
transfer to the lattice via the electron-phonon coupling.  If the
lattice can transfer its energy effectively enough to the underlying
substrate, a hot-electron situation may be reached, where the
electronic temperature $T_e$ is considerably higher than that of the
(acoustic) phonons, $T_{ac}$.
\cite{allen87,wellstood94,giazotto06} More
generally, such a situation with well defined separate electron and
phonon temperatures is known as quasiequilibrium.  In analyzing
experiments or devices that involve heating (or
cooling\cite{hekking08}) effects of this kind, it is essential to be
able to model the heat currents between the various
subsystems.\cite{giazotto06} This is particularly important in
two-dimensional graphene, which has a small thermal volume and is thus
easily overheated.  Of central interest here is the heat current
between the electrons and the phonons, on which we concentrate in this
paper.

In metals at sufficiently low temperatures only acoustic
phonons are relevant, and of those the longitudinal (LA) modes have the
strongest coupling constants.  At very low temperature the power
transferred between the electrons and LA
phonons (assuming quasiequilibrium) is typically of the form
\begin{equation} \label{e.ltform}
Q_{\rm e-ac}=V_d\Sigma
(T_e^\delta-T_{ac}^\delta),
\end{equation}
where $V_d$ is the $d$-dimensional volume and $\Sigma$ is a coupling
constant. This form is well known in the case of simple
three-dimensional (3-D) metals, where the exponent is $\delta=5$
(Refs.\ \onlinecite{wellstood94,Mahan3rd}).  In lower-dimensional and
disordered systems different exponents have been found. In disordered
thin films $\delta=6$ has been observed,\cite{sergeev00,karvonen05}
and in thin doubly clamped metallic beam, where the electrons remain
3-D but vibrations are one-dimensional, $\delta=3$ and $V_d$ is the
length of the beam.\cite{hekking08}

The electron cooling power and temperature relaxation times were recently
studied theoretically also for monolayer graphene
(MLG).\cite{kubakaddi09,bistritzer09,tsedassarma09} Graphene is a
single sheet of graphite, \emph{i.e.}\ a two-dimensional (2-D)
honeycomb lattice of carbon, which has (semi)metallic
properties.\cite{novoselov04,castroneto09} In its monolayer form the valence and
conduction bands touch at the two K points of the
Brillouin zone (or Dirac points), around which the electron dispersion
relation is conical.  In bilayer graphene (BLG) the dispersion is approximately
parabolic.\cite{mccann06} The most important difference of graphene to
other metals or semiconductors is the ``chiral'' character of the charge
carriers.  However,
close to the charge neutrality point the properties of graphene differ
from other metals already by the special forms of the dispersion
relations.  In Ref.\ \onlinecite{kubakaddi09} it was found that at low
temperature the power transferred to in-plane LA phonons in MLG is of the
usual form (\ref{e.ltform}) with $\delta=4$ [see Eq.\ (\ref{e.mls})
below].  In Ref.\ \onlinecite{bistritzer09}, on the other hand, it
was shown that in the neutral (or high-temperature) regime of MLG the
power has a form asymmetric in $T_e$ and $T_{ac}$,
\begin{equation}\label{e.htform}
Q_{\rm e-ac}=V_dg(\mu,T_e)(T_e-T_{ac}),
\end{equation}
where $V_d=A$, the area of the sample,
$g(\mu,T_e)$ is a function which we specify in Eq.~\eqref{e.mlf} below,
and $\mu$ the chemical potential measured from the 
Dirac point.
 
In this paper we revisit the problem of electron-phonon power transfer
in MLG and consider also the case of BLG 
in the parabolic-band approximation.  We show 
that the conditions for the validity of the ``low-temperature'' and 
``high-temperature'' results for MLG mentioned above are roughly
$T_e,T_{ac} \ll T_{\rm BG,MLG}$ and $T_e, T_{ac} \gg T_{\rm BG,MLG}$,
respectively, where $T_{\rm BG,MLG}=2(c/v)|\mu|/k_B$ is the 
Bloch-Gr\"uneisen temperature of MLG.
Here $v\approx 1\times 10^6$ m/s
is the Fermi speed in MLG and $c\approx 2 \times 10^4$ m/s $\ll v$ is
the speed of sound. 
These two limits have been termed the Bloch-Gr\"uneisen (BG) 
and equipartition (EP) regimes, respectively.\cite{hwang08}
In these limits we recover the previously
derived results [Eqs.\ (\ref{e.mls}) and (\ref{e.mlf})],  
but in addition to studying the limits analytically, we evaluate 
the power numerically in the crossover region.
We then repeat the same analysis for BLG, where
$T_{\rm BG,BLG}=2(c/v)\sqrt{\gamma_1|\mu|}/k_B$, with $\gamma_1$
the interlayer coupling amplitude.  
Also for BLG, in the BG regime the usual
power-law form (\ref{e.ltform}) for acoustic phonons is found,
with $\delta=4$, while the result in the opposite regime
has the form (\ref{e.htform}) [Eqs.\ (\ref{e.bls}) 
and (\ref{e.blf})].  
In fact the exponent $\delta=4$ is independent of the
chirality of the carriers, since it appears that in non-disordered
systems at low temperatures quite generally $\delta=d+2$, where $d$ is
the smaller of the dimensions of the electron and the phonon
systems. However, the coupling constants $\Sigma$ for MLG and BLG
depend rather differently on doping, with $\Sigma\propto|\mu|$ in MLG, and
$\Sigma\propto 1/\sqrt{|\mu|}$ in BLG.  
These differences in $\Sigma$ can be understood based on 
the electron-phonon relaxation rates and electronic 
specific heats.
In the EP regime we find that the
temperature dependence of  $g(\mu,T_e)$ is weaker in
BLG ($g\sim T_e$) than in MLG ($g\sim T_e^4$).

At very low temperatures the most efficient energy relaxation
mechanism for electrons is diffusion. In order to measure $Q_{e-ac}$
for determining the coupling constant $\Sigma$, the temperature should
be higher than a certain value which we estimate in
Sec.\ \ref{s.discussion} for a typical experimental situation.  On the
other hand, at even higher temperatures the optical phonon power
becomes dominant. To estimate also the temperature for this crossover,
we consider some simplified models for optical phonons that may be
relevant. The most obvious ones are the intrinsic in-plane
longitudinal (LO) or transverse (TO) modes.  For graphene on a
dielectric substrate, the surface optical phonons of the dielectric
must also be considered.\cite{fratini08,meric08} We find that the
latter can begin to dominate the energy relaxation already at much
lower temperatures than the intrinsic phonons.

In suspended graphene, in addition to the LA phonons, also
out-of-plane acoustic (flexural) modes should be
considered.\cite{mariani08} However, these are disregarded here as
their coupling to electrons is only of second order in the
displacements and since they should not be important for graphene on a
substrate. On the other hand, we also do not discuss explicitly the
coupling of the graphene phonons to those of the substrate, or the
heat transfer (Kapitza resistance) between them,\cite{chen09} which,
although an important part of the heat-balance problem, depends on the
details of the interface and is difficult to model microscopically.
This coupling may to some extent affect the acoustic phonon
dispersions, and the relevant LA phonons should possibly be understood
as collective surface acoustic modes of the coupled graphene-substrate
system.  Finally, we do not take into account effects of impurities,
or discuss explicitly electron-electron interactions, which are
assumed to be strong enough to keep the electrons in quasiequilibrium.

This paper is organized as follows. In Sec.\ \ref{s.boltzmann} we
sketch our Boltzmann-Golden Rule approach for calculating the heat
current and briefly discuss the quasiequilibrium approximation.
Next we give a brief reminder on the low-energy electronic structure of graphene 
in Sec.\ \ref{s.bands} and then continue to 
consider the coupling of the electrons to  LA phonons via 
a deformation-potential approach in Secs.\ \ref{s.eac} and \ref{s.acres}.  In
Sec.\ \ref{s.eop}, we describe our models for the optical
phonons and their coupling to electrons, and the calculation of the
corresponding heat currents. Finally, we discuss the experimental 
consequences of our results and the crossover temperatures in Sec.\ \ref{s.discussion}.


\section{Electron-phonon heat
transfer from the Boltzmann theory and Fermi Golden Rule} \label{s.boltzmann}

Below we derive a general expression for the electron-phonon heat
current by employing a Boltzmann collision integral in the Fermi
Golden Rule approximation. In what follows, we disregard the spin and
valley indices, which only appear as an additional degeneracy factor
$g_e=4$ in the results.

The position-independent Boltzmann equation describing the occupation
probability $f_{\vec{k}}^\alpha$ of the electron excitation with
momentum $\hbar\vec{k}$ in band $\alpha$ is
\begin{equation}
\partial_tf_{\vec{k}}^\alpha
= S_{e-ph}(f_{\vec{k}}^\alpha).
\end{equation}
Here the collision integral is given by
\begin{equation}\begin{split}
S_{e-ph}(f_{\vec{k}}^\alpha) =& -\sum_{\vec{p}\beta}
[
f_{\vec{k}}^\alpha(1-f_{\vec{p}}^\beta) 
W_{\vec{k}\alpha\rightarrow\vec{p}\beta} \\ 
&-f_{\vec{p}}^\beta(1-f_{\vec{k}}^\alpha) 
W_{\vec{p}\beta\rightarrow\vec{k}\alpha} 
],
\end{split}\end{equation}
where the Golden-Rule scattering rates are 
\begin{equation}\begin{split}
W_{\vec{k}\alpha\rightarrow\vec{p}\beta} 
=&
\frac{2\pi}{\hbar}
\sum_{\vec{q}\gamma}
w_{\vec{k}\vec{p},\vec{q}}^{\alpha\beta,\gamma}
[
(n_{\vec{q}}^\gamma+1)
\delta_{\vec{k},\vec{p}+\vec{q}}
\delta(\epsilon_{\vec{k}\vec{p}}^{\alpha\beta}-\omega_{\vec{q}}^\gamma) \\
&+
n_{\vec{q}}^\gamma
\delta_{\vec{k},\vec{p}-\vec{q}}
\delta(\epsilon_{\vec{k}\vec{p}}^{\alpha\beta}+\omega_{\vec{q}}^\gamma)
].
\end{split}\end{equation}
In these $n_{\vec{q}}^\gamma$ is the distribution function of phonons
with momentum $\hbar\vec{q}$ and band index $\gamma$, 
$\epsilon_{\vec{k}}^\alpha$ and $\omega_{\vec{q}}^\gamma$ are the 
electron and phonon excitation energies, and we defined 
$\epsilon_{\vec{k}\vec{p}}^{\alpha\beta}$
$=\epsilon_{\vec{k}}^{\alpha}-\epsilon_{\vec{p}}^{\beta}$.
We assume the electron-phonon coupling constants to 
satisfy the symmetries 
$w_{\vec{k}\vec{p},\vec{q}}^{\alpha\beta,\gamma}
=w_{\vec{p}\vec{k},\vec{q}}^{\beta\alpha,\gamma}
=w_{\vec{k}\vec{p},-\vec{q}}^{\alpha\beta,\gamma}$.
The power with which the phonons cool the electrons 
(\emph{i.e.}\ the electron-phonon heat current)
is defined by 
\begin{equation}\begin{split}
Q & = 
-\partial_t
\sum_{\vec{k}\alpha}
\epsilon_{\vec{k}}^\alpha
f_{\vec{k}}^\alpha
=
-\sum_{\vec{k}\alpha}
\epsilon_{\vec{k}}^\alpha
S_{e-ph}(f_{\vec{k}}^\alpha).
\end{split}\end{equation}
It is convenient to divide the power into two terms according to
whether they describe induced (stimulated) or
spontaneous processes.\cite{wellstood94} Thus we find
\begin{equation}\begin{split}
Q = Q_{\rm ind} + Q_{\rm spont}, 
\end{split}\end{equation}
where
\begin{subequations}
\begin{align}\begin{split}
Q_{\rm ind} = & +\frac{2\pi}{\hbar} 
\sum_{\vec{q}\gamma}
\sum_{\vec{p}\beta} 
\sum_{\vec{k}\alpha} 
\epsilon_{\vec{k}\vec{p}}^{\alpha\beta}
w_{\vec{k}\vec{p},\vec{q}}^{\alpha\beta,\gamma}
[f(\epsilon_{\vec{k}}^\alpha) - f(\epsilon_{\vec{p}}^\beta)] \\
&\times n_{ph}(\omega_{\vec{q}}^\gamma)
\delta_{\vec{k},\vec{p}+\vec{q}}
\delta(\epsilon_{\vec{k}\vec{p}}^{\alpha\beta}-\omega_{\vec{q}}^\gamma),
\end{split}\\
\begin{split}
Q_{\rm spont} = & -\frac{2\pi}{\hbar} 
\sum_{\vec{q}\gamma}
\sum_{\vec{p}\beta} 
\sum_{\vec{k}\alpha} 
\epsilon_{\vec{k}\vec{p}}^{\alpha\beta}
w_{\vec{k}\vec{p},\vec{q}}^{\alpha\beta,\gamma}
[f(\epsilon_{\vec{k}}^\alpha) - f(\epsilon_{\vec{p}}^\beta)] \\
&\times n_{e}(\omega_{\vec{q}}^\gamma)
\delta_{\vec{k},\vec{p}+\vec{q}}
\delta(\epsilon_{\vec{k}\vec{p}}^{\alpha\beta}-\omega_{\vec{q}}^\gamma).\label{e.qspont}
\end{split}\end{align}
\end{subequations}
Here and in the following we assume that the system is in
quasiequilibrium where electrons are described by the electron
temperature $T_e$ and chemical potential $\mu$, and phonons by the
temperature $T_{ph}$.  Thus
$f_{\vec{k}}^\alpha=f(\epsilon_{\vec{k}}^\alpha)$ and
$n_{\vec{q}}^\gamma=n_{ph}(\omega_{\vec{q}}^\gamma)$ where
$f(E)=\{\exp[(E-\mu)/k_BT_e]+1\}^{-1}$ is the Fermi function and
$n_{i}(\omega)=[\exp(\omega/kT_i)-1]^{-1}$ the Bose function at
temperature $T_i$.  
To arrive at the form \eqref{e.qspont} we have applied
$f(\epsilon_{\vec{k}}^\alpha)[1-f(\epsilon_{\vec{p}}^\beta)]=
-[f(\epsilon_{\vec{k}}^\alpha)-f(\epsilon_{\vec{p}}^\beta)]
n_e(\epsilon_{\vec{k}\vec{p}}^{\alpha\beta})$, which is
valid in equilibrium.

The only difference between
$Q_{\rm spont}$ and $Q_{\rm ind}$
is therefore the sign and the temperature that enters the Bose function.
Writing
$Q_{\rm ind} =Q_{\rm ind}(\mu,T_e,T_{ph})$,
we have 
\begin{equation}\begin{split} \label{e.qsymm}
Q_{\rm spont}(\mu,T_e)& =-Q_{\rm ind}(\mu,T_e,T_e).
\end{split}\end{equation}
Due to this symmetry, it is only necessary to calculate 
$Q_{\rm ind}$ explicitly.  It is easy to see that the net 
power transfer is zero if $T_e=T_{ph}$.

Besides the heat current $Q$, another useful quantity is the 
linear-response thermal conductance, defined as
\begin{equation}\begin{split}
G(\mu,T) = \frac{\partial}{\partial \Delta T}
Q\left(\mu,T+\frac{\Delta T}{2},T-\frac{\Delta T}{2}\right)\bigg|_{\Delta T=0},
\end{split}\end{equation}
where $Q=Q(\mu,T_e,T_{ac})$ and 
$\Delta T=T_e-T_{ac}$, $T=(T_e+T_{ac})/2$.
This is more convenient for characterizing the strength of the 
electron-phonon coupling, because it only depends on a single
temperature.

A more physical reason for concentrating on the linear-response regime
comes from the assumption of quasiequilibrium.  The latter is a
commonly used approximation that avoids the need to include the
electron-electron collision integral and an equally detailed treatment
of the phonon kinetics.\cite{giazotto06} Electrons remain in internal
equilibrium if the electron-electron interactions are sufficiently
strong.\cite{bistritzer09b} (For an example of a case where this is
not valid, see Ref.\ \onlinecite{kabanov08}.)  The same applies to
phonons if their relaxation is strong enough, which is typically the
case for systems on a substrate. However, although these conditions
might not always be satisfied far from equilibrium, we do not expect
them to be a serious concern in the linear response regime.


\section{Electronic structure of
monolayer and bilayer graphene} \label{s.bands}

\begin{figure}[!tb] 
\includegraphics[width=0.9\linewidth,clip=]{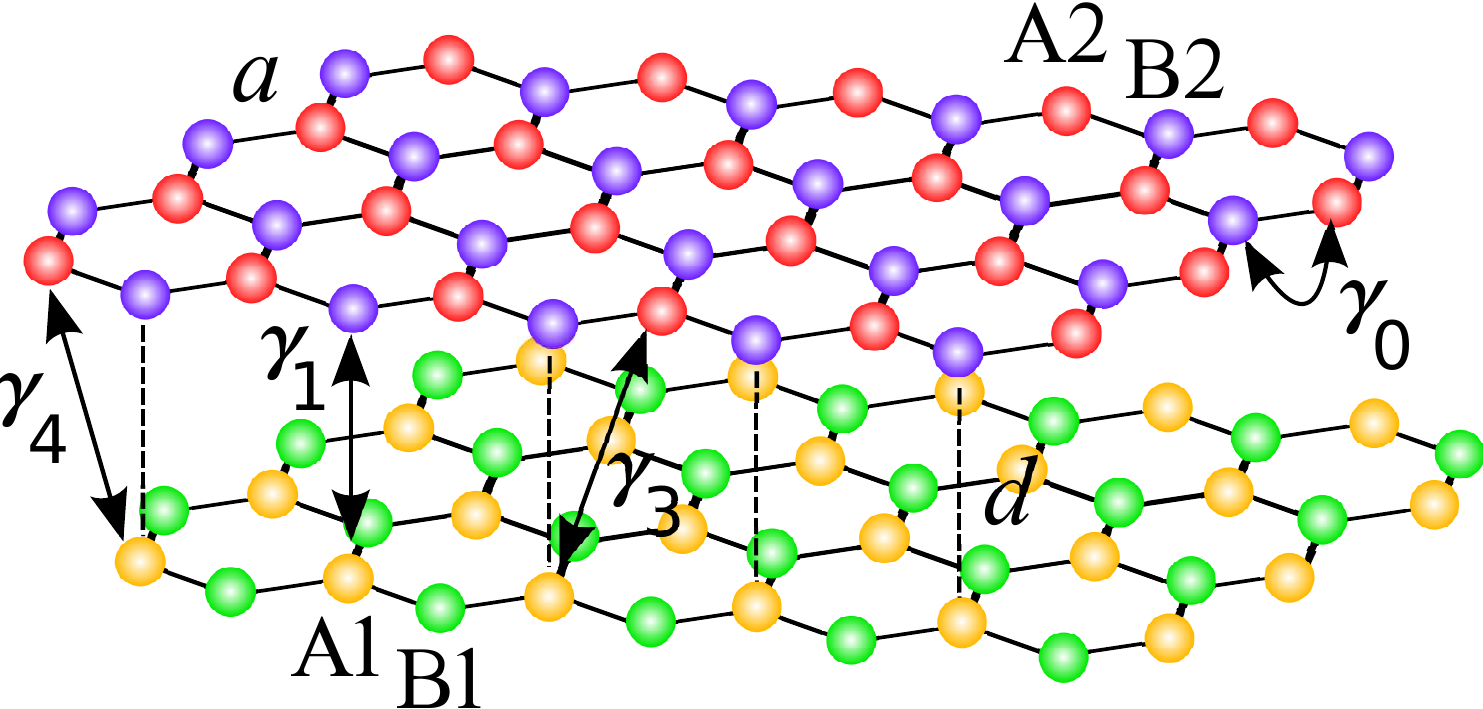}
\caption{(Color online) Geometrical structure of bilayer graphene 
  with AB, or Bernal, stacking and the
  Slonczewski-Weiss-McClure band parameters $\gamma_{0,1,3,4}$.  The
  A1, B1, A2, and B2 atoms of the two sublattices and layers are indicated. 
  The nearest-neighbor distance is $a\approx 0.14$ nm,
  and the interlayer distance is $d\approx 0.37$ nm.}
\label{f.bilayer_struct}
\end{figure}

Here we give a very brief reminder on the low-energy band structure of
graphene. Figure \ref{f.bilayer_struct} shows the geometrical
structure of BLG. For a tight-binding description we employ the
Slonczewski-Weiss-McClure
parameters.\cite{slonczewski58,mcclure57,castroneto09,dresselhaus02}
In what follows we only use the parameters $\gamma_0\approx 3$ eV,
$\gamma_1\approx 0.4$ eV and set $\gamma_3=\gamma_4=0$, which is a
reasonable approximation at not-too-low energies.\cite{mccann06} 
MLG is obtained by setting also $\gamma_1=0$ and concentrating on one 
of the layers. The unit cell then has two atoms and the 
Hamiltonian for each of the K point of the reciprocal space 
is a $2\times 2$ matrix, whose eigenvalues give the conical ``Dirac'' spectrum. 
In the case of BLG the unit cell has four atoms, and thus the low-energy
Hamiltonian is a $4\times 4$ matrix.
However, using perturbation theory to leading order in $1/\gamma_1$, 
it may be replaced with an effective
$2\times 2$ Hamiltonian in the basis of the ``uncoupled'' atoms A2 and
B1.\cite{mccann06,castroneto09,nilsson08} The resulting dispersion
relation is parabolic.

Using this parabolic-band approximation for BLG, the dispersion relations
of the valence ($\mathrm{v}$) and conductance ($\mathrm{c}$) bands 
and the corresponding 
eigenstates may be written collectively as
\begin{equation}\begin{split} \label{e.blspinor}
\epsilon_{\vec{k}}^{\alpha}
&= 
s_\alpha(\hbar v k)^n/\gamma_1^{n-1}, \\
\phi_{\vec{k}}^{\alpha} & = 
\frac{1}{\sqrt{2}}
[1, \pm s_\alpha e^{\pm n i \phi_{\vec{k}}}]^T,
\end{split}\end{equation}
where 
$n=1$ for MLG, and $n=2$ for BLG,
$s_\alpha=\pm 1$ for $\alpha={\rm c,v}$,
$\phi_{\vec{k}}=\arctan(k_y/k_x)$,
$k=|\vec{k}|$,
and $\hbar v=3a\gamma_0/2$, with $v\approx 10^6$ m/s.
The upper and lower signs correspond to the 
two different K points (valleys), relative to 
which the wave vectors $\vec{k}$ are counted.
The zero of energy is set at the charge neutrality point.

These low-energy results are sufficient for the
discussion of interactions of the electrons with acoustic phonons. 
Atomic-scale scatterers such as short-wavelength optical phonons 
can also couple the two K points,\cite{suzuura08,rana09} 
but for simplicity we disregard intervalley scattering in our 
discussion of optical phonons below.


\section{Acoustic phonons in monolayer and bilayer graphene} \label{s.eac}

Long-wavelength acoustic phonons may be treated with continuum
theories to a good approximation. An atomistic treatment is required
only for optical phonons, see Sec.\ \ref{s.eop}.  Here we do not
discuss the actual calculation of the phonon modes, which could be
described in terms of elasticity theory.\cite{suzuura02,mariani08,castroneto09}


\subsection{Coupling via deformation potential} \label{s.defpot}

The dominant form of electron-phonon coupling for long-wavelength
acoustic modes is due to the deformation
potential.\cite{suzuura02,mariani08,suzuura08} For in-plane modes the potential
is of the form $D\nabla\cdot\vec{u}$, where $\vec{u}(\vec{r})$ is the
displacement vector from elasticity theory, and $D$ is the coupling
constant, for which values in the range $D=10-50$ eV have been used
(Refs.\ \onlinecite{suzuura02,kubakaddi09,bistritzer09}).

The deformation potential is nonzero only for longitudinal (LA) modes.  
Thus we may use the expansion in terms of longitudinal plane waves:
$\vec{u}(\vec{r}) = \sum_{\vec{q}} u_{\vec{q}} \hat{\vec{q}}
e^{i\vec{q}\cdot\vec{r}}.$
Quantizing this with 
$u_{\vec{q}}=i\sqrt{\hbar^2/(2M\omega_{\vec{q}})}(b_{\vec{q}}+b_{-\vec{q}}^\dagger)$,
where $b^{\dagger}_{\vec{q}}$ and $b_{\vec{q}}$ are the
phonon creation and annihilation operators, 
the electron-phonon coupling Hamiltonian becomes
\begin{equation}\begin{split}
\hat{H}_{\rm e-ac} 
=& \sum_{\vec{k}\alpha,\vec{p}\beta}\sum_{\vec{q}}
M_{\vec{p}\vec{k},\vec{q}}^{\beta\alpha}
c_{\vec{p}\beta}^\dagger c_{\vec{k}\alpha}
(b_{\vec{q}} + b_{-\vec{q}}^\dagger),
\end{split}\end{equation}
with
$M_{\vec{p}\vec{k},\vec{q}}^{\beta\alpha}=- \sqrt{\hbar^2/(2M\omega_{\vec{q}})} Dq
\langle \vec{p}\beta|e^{i\vec{q}\cdot\vec{r}}|\vec{k}\alpha\rangle$. 
Here $\omega_{\vec{q}}=\hbar c q$ is the LA phonon dispersion and
$M=A\rho$ is the total mass, with $\rho$ the mass density (of MLG or BLG)
and $A$ the
area of the system.  The wave functions are
$\langle\vec{r}|\vec{k}\alpha\rangle=e^{i\vec{k}\cdot\vec{r}}\phi_{\vec{k}}^\alpha/\sqrt{A}$, where 
Eqs.\ (\ref{e.blspinor}) should be used for $\phi_{\vec{k}}^\alpha$.
The coupling constants are then identified from
$|M_{\vec{p}\vec{k},\vec{q}}^{\beta\alpha}|^2=w_{\vec{k}\vec{p},\vec{q}}^{\alpha\beta}
\delta_{\vec{p},\vec{k}+\vec{q}}$, which yields\cite{tsedassarma09,rotkin09}
\begin{equation}\begin{split} \label{e.gaccoupling}
w_{\vec{k}\vec{p},\vec{q}}^{\alpha\beta}
= &
\frac{\hbar^2}{2M\omega_{\vec{q}}} D^2q^2
F_{\alpha\beta}(\theta).
\end{split}\end{equation}
Here $F_{\alpha\beta}(\theta)=(1+s_\alpha s_\beta\cos n\theta)/2$,
$\theta=\phi_{\vec{p}}-\phi_{\vec{k}}$, 
and $n=1,2$ for MLG and BLG, respectively.  This shows that for MLG
backscattering from acoustic phonons is weak, whereas for BLG
there is no difference between forward and backward (intraband)
scattering.


\subsection{Conservation laws} \label{s.conserv}

In scattering of electrons from acoustic phonons, the final wave
vectors $\vec{p}$ allowed by energy and momentum conservation laws
(for given initial $\vec{k}$) may be determined from the equation
$\epsilon_{\vec{p}}^\beta = \epsilon_{\vec{k}}^\alpha + s\hbar c |\vec{p}-\vec{k}|$,
where $s=+1$ for absorption and $s=-1$ for emission, and
$\epsilon_{\vec{k}}^\alpha$ and $\epsilon_{\vec{p}}^\beta$ are the
initial and final energies, respectively.  The
right-hand side describes a conical surface in momentum-energy space
with the apex at $(\vec{k},\epsilon_k^\alpha)$ and the
left-hand side is another surface. In the case of MLG this surface is
also a cone, while for BLG in the parabolic-band approximation it is a
paraboloid.  The allowed $\vec{p}$ values now lie on the
$\vec{p}$-plane projection of the curve defined by the intersection of
the two surfaces.  These may be worked out analytically, and they are
needed for the numerical evaluation of the heat current. 
The details are given in App.\ \ref{s.anasol}.  It is noteworthy that in the
case of MLG there cannot be any interband scattering
($\beta\neq\alpha$) when the phonon dispersion is
linear,\cite{tsedassarma09} while for BLG interband scattering is
possible if the initial $k$ is small enough.  However, in practice
this scattering is strongly suppressed by the coupling constant
(\ref{e.gaccoupling}).


\section{Heat current between electrons and acoustic phonons} \label{s.acres}

Here we use the results of Sec.\ \ref{s.boltzmann} for graphene,
with the LA phonon coupling constant given by Eq.\ (\ref{e.gaccoupling}).
First we present the general result which can be evaluated numerically,
and then discuss analytic approximations in the BG and EP limits.


\subsection{Numerical solution}

\begin{figure}[!tb] 
\includegraphics[width=0.9\linewidth,clip=]{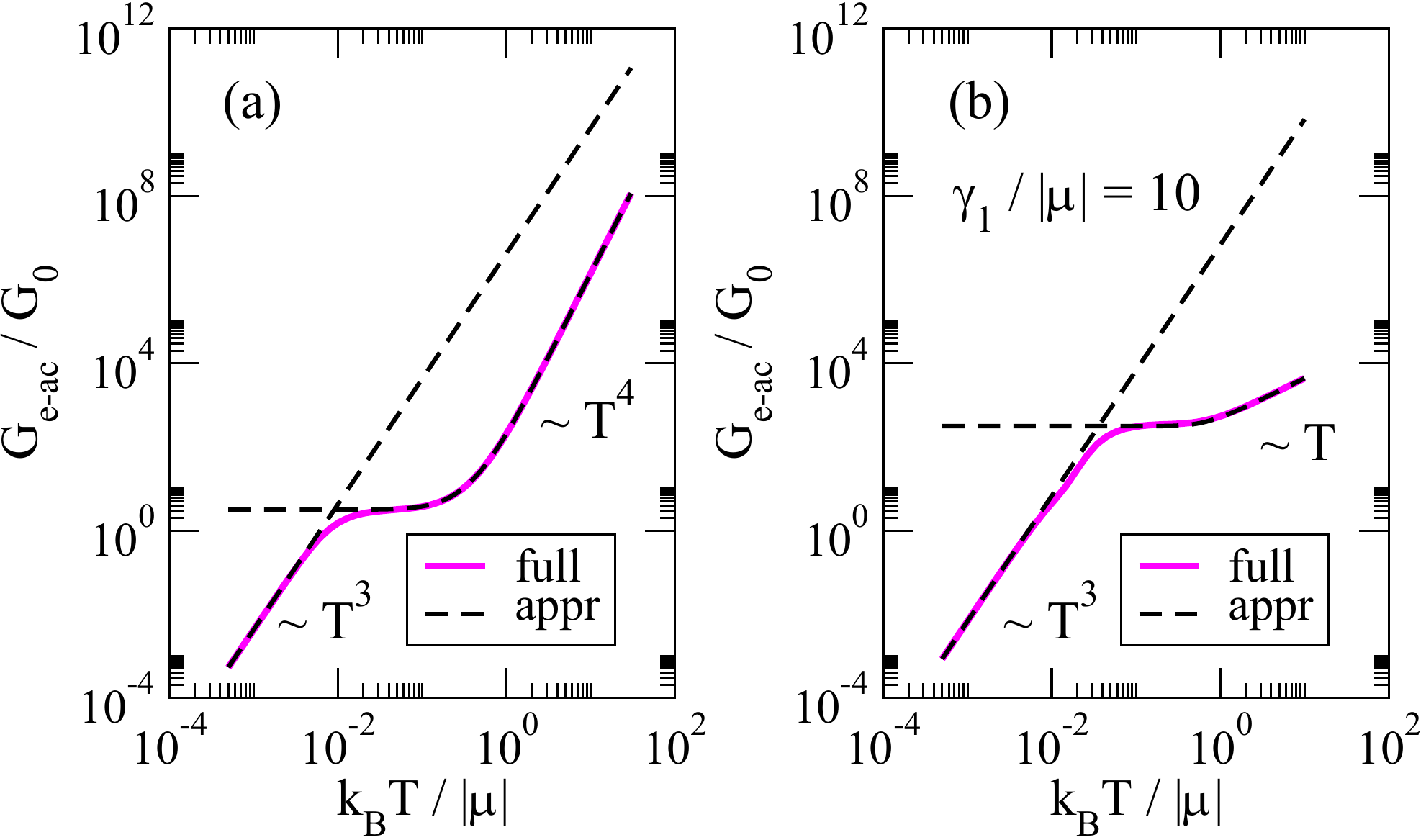}
\caption{(Color online) Thermal conductance $G_{e-ac}(\mu,T)$ for MLG
  (a) and BLG (b).  The unit is
  $G_0=g_eAD^2\hbar|\mu|^4k_B/[8\pi^2\rho (\hbar v)^6]$ for MLG and
  $G_0=g_eAD^2\hbar\gamma_1|\mu|^3k_B/[16\pi^2\rho(\hbar v)^6]$ for
  BLG.  The parameters are $c/v=0.02$ and $\gamma_1/|\mu|=10$ in
  (b). The solid lines (magenta) indicate the full numerical solutions, 
  and the dashed lines (black) the analytical approximations.  }
\label{f.tcondplots}
\end{figure}

For acoustic LA phonons the most general expression
for $Q_{\rm ind}$ without any approximations is
\begin{equation}\begin{split} \label{e.qindgene}
Q_{\rm ind} = &-\frac{g_eA\hbar D^2}{2(2\pi)^2\rho} 
\sum_{\alpha\beta} 
\int_0^\infty dk k \int_{q_{\rm min}^{\alpha\beta}}^{q_{\rm max}^{\alpha\beta}} dq q^2
\\
&\times
\frac{1+s_\alpha s_\beta y(k,q)}{\sqrt{1-(x(k,q))^2}} 
|\hbar v_k^\alpha [p(k,q)/k]^{n-2}|^{-1}
\\
&\times
n_{ac}(\hbar c q)
[f(\epsilon_k^\alpha-\hbar c q)-f(\epsilon_k^\alpha)],
\end{split}\end{equation}
with $n=1,2$ as above. 
Here
$v_k^\alpha=s_\alpha n(\hbar v)^n(k/\gamma_1)^{n-1}/\hbar$ is the Fermi speed
and $x(k,q) =-\{[p(k,q)]^2 - k^2 - q^2\}/(2kq)$
is the cosine of the angle between the incoming $\vec{k}$
and the phonon $\vec{q}$. In the latter 
$p(k,q)=(s_\alpha s_\beta - s_\beta\hbar c q \gamma_1^{n-1}/(\hbar v k)^n)^{1/n}k$
is the length of
$\vec{p}=\vec{k}-\vec{q}$ after imposing conservation laws. Moreover,
$y(k,q) = n(z(k,q))^n-n+1$,
where $z(k,q) = [k - q x(k,q)]/p(k,q)$
is the cosine of the angle between $\vec{k}$ and $\vec{p}$.
Finally, the correct limits $q_{\rm max,min}^{\alpha\beta}$ of the $q$
integral are obtained from the conservation laws, as discussed in Sec.\ \ref{s.conserv}.
Using these results together with Eq.\ (\ref{e.qsymm}),
the total power $Q_{e-ac}$ may be obtained numerically.
An example of the corresponding thermal
conductance $G_{e-ac}(\mu,T)$ is represented by the solid lines in
Fig.\ \ref{f.tcondplots}. 

To obtain analytic estimates, we first expand Eq.\ (\ref{e.qindgene}) 
to leading order in $c/v \ll 1$.  In this approximation we may
neglect interband transitions ($\beta\neq\alpha$) also for BLG 
and use $x(k,q)= q/2k$, $p(k,q) = k$, $q_{\rm max}^{\alpha\alpha}=2k$, 
and $q_{\rm min}^{\alpha\alpha}=0$.
In the BG and EP limits
the forms of Eqs.\ (\ref{e.ltform}) and (\ref{e.htform}) are then recovered.
Below we give the results for the coefficients $\Sigma$  and $g(\mu,T_e)$ 
for MLG and BLG.  
The thermal conductances $G_{e-ac}$ obtained from them 
are shown as the dashed lines in Fig.\ \ref{f.tcondplots}. 
The agreement with the full numerical solution is very good.
The most notable difference between MLG and BLG 
is the much weaker temperature dependence
of $G_{e-ac}$ for $k_BT_e\gg|\mu|$ in BLG.


\subsection{Limits for monolayer graphene}

First consider the case of MLG ($n=1$).
For a low-temperature approximation of Eq.\ (\ref{e.qindgene}) we may estimate
$\epsilon_k^\alpha=s_\alpha\hbar vk\sim\mu$ and $\hbar c q \sim k_B
T_{ac}$. If $k_BT_e,k_BT_{ac}\ll 2(c/v)|\mu|=k_BT_{\rm BG,MLG}$, 
we may set $q/2k\rightarrow 0$ in the $q$ integral and extend the
upper limit $q_{max}=2k\rightarrow\infty$.  Since $\hbar c
q\ll|\epsilon_k^\alpha|$, we expand the Fermi function to the
first order in $\hbar c q$.  The $T_e$ dependence then drops out of
$Q_{\rm ind}$ so that the total power takes on the symmetric
form of Eq.\ (\ref{e.ltform}), where $V_d=A$, $\delta=4$, and
\begin{equation} \label{e.mls}
\Sigma=\frac{\pi^2D^2|\mu|k_B^4}{15\rho \hbar^5v^3c^3}.
\end{equation}
The same result was derived previously in Ref.\ \onlinecite{kubakaddi09}.

The coupling constant $\Sigma$ may be expressed in terms of the 
electron-phonon relaxation rate at the Fermi level and the electronic 
specific heat.\cite{wellstood94} 
We start from the definition of the relaxation rate
$\tau_{\vec{k}\alpha}^{-1}=-\partial_{f_{\vec{k}}^\alpha}S_{e-ph}(f_{\vec{k}}^\alpha)|_{f=f_0}$,
where $f_0(E)=\{\exp[(E-\mu)/k_BT]+1\}^{-1}$ and $T$ is the lattice temperature.
A similar calculation as for $Q_{\rm ind}$ yields
\begin{equation} \label{e.mlratelt}
\tau_{k_F}^{-1}=\frac{\pi D^2 k_B^2}{4\rho\hbar^3vc^3}T^2,
\end{equation}
where $k_F=|\mu|/(\hbar v)$.
(We note that unlike this result, the 
transport relaxation rate\cite{hwang08,mariani08} is $\propto T^4$.)
The electronic specific heat 
is given by $C=(\pi^2/3)\nu(\mu)k_B^2T$, where the density of 
states including the degeneracy $g_e=4$
is $\nu(E)=2|E|/(\pi\hbar^2v^2)$. Identifying
$\tau_{k_F}^{-1}=\alpha^*T^2$ and $C=\gamma^* T$,
we find $\Sigma=(2/5)\alpha^*\gamma^*$.

Seeking a high-temperature approximation, we require
$\hbar c q \ll k_B T_{e}, k_B T_{ac}$, which allows us to expand
also the Bose function.  Since for a given
$k$ the maximal $q$ is $2k$, and the maximal relevant $k$ is
determined by $|\epsilon_k^\alpha|\sim \mathrm{max}(|\mu|,k_BT_e)$,
the limits translate to $2(c/v)\mathrm{max}(|\mu|,k_BT_e)\ll k_B T_{e},k_BT_{ac}$.  
This is the case considered in Ref.\ \onlinecite{bistritzer09},
and since $c/v\ll 1$ the limit is equivalent to 
$T_e,T_{ac}\gg T_{\rm BG,MLG}$, if $T_e\sim T_{ac}$.
In this limit, $Q_{\rm ind}$ depends on both $T_e$
and $T_{ac}$. Using Eq.\ (\ref{e.qsymm})
then leads to a total power of the asymmetric form in Eq.\ (\ref{e.htform}),
where $V_d=A$ and
\begin{equation}\begin{split} \label{e.mlf}
g(\mu,T_e)
=&\frac{D^2 k_B}{30\pi\rho\hbar^5 v^6} 
[15\mu^4 + 30\pi^2\mu^2(k_BT_e)^2 \\
&+ 7\pi^4(k_BT_e)^4].
\end{split}\end{equation}
For $k_{\rm B}T_{\rm BG,MLG}\ll k_BT_e\ll|\mu|$ this yields $g(\mu,T_e)\propto
\mu^4$, and in the limit $|\mu|\ll k_BT_e$ we find
$g(\mu,T_e)\propto T_e^4$.

For completeness we mention also the result\cite{hwang08} 
for the high-temperature relaxation rate in MLG:
\begin{equation}\begin{split} \label{e.mlrateht}
\tau_{k_F}^{-1}=\frac{D^2|\mu| k_B}{2\rho\hbar^3v^2c^2}T.
\end{split}\end{equation}
In this case the relation to $Q_{e-ac}$ is more complicated 
than at low temperature.


\subsection{Limits for bilayer graphene}

Next consider BLG ($n=2$).
In this case, besides $\mu$, $T_e$, and $T_{ac}$,
there is an additional energy scale determined by $\gamma_1$.  In order for the parabolic
two-band approximation to be valid, we must require $\mathrm{max}(|\mu|,k_BT_e)\ll \gamma_1$.

To find a low-temperature approximation, we follow similar steps as for MLG.
The assumed limit is now
$k_BT_e,k_BT_{ac}\ll 2(c/v)\sqrt{\gamma_1|\mu|}=k_BT_{\rm BG,BLG}$
(with $|\mu|\ll\gamma_1$), using which we find 
Eq.\ (\ref{e.ltform}) with $\delta=4$, but this time
\begin{equation} \label{e.bls}
\Sigma=\frac{\pi^2 D^2\gamma_1k_B^4}{60\rho \hbar^5v^3c^3}
\sqrt{\frac{\gamma_1}{|\mu|}}.
\end{equation}
The scaling with temperature is thus similar to the MLG
case, but the $\mu$ dependence is different.
Result (\ref{e.bls}) is valid also in a normal 
2-D system with effective mass $\gamma_1/(2v^2)$.

For BLG we find the relaxation rate
\begin{equation} \label{e.blrate}
\tau_{k_F}^{-1}=\frac{\pi D^2 k_B^2}{8\rho\hbar^3vc^3}
\sqrt{\frac{\gamma_1}{|\mu|}}T^2,
\end{equation}
where $k_F=\sqrt{\gamma_1|\mu|}/(\hbar v)$.
(The corresponding transport relaxation rate is again 
$\propto T^4$.) Now the density of 
states entering $C$ is approximately 
$\nu(E)\approx \gamma_1/(\pi\hbar^2v^2)$.
Also in this case the definitions 
$\tau_{k_F}^{-1}=\alpha^*T^2$ and $C=\gamma^* T$ lead 
to $\Sigma=(2/5)\alpha^*\gamma^*$.

For BLG the high-temperature (EP) limit 
$\hbar c q \ll k_B T_{e},k_BT_{ac}$ may be written as
$2(c/v)\sqrt{\gamma_1\mathrm{max}(|\mu|,k_BT_e)}\ll k_B T_{e},k_BT_{ac}$.
This is again equivalent to 
$T_{e},T_{ac}\gg T_{BG,BLG}$, if additionally
$T_e\sim T_{ac}\gg 2(c/v)^2\gamma_1/k_{\rm B}$.
The total power again takes the form of Eq.\ (\ref{e.htform}), 
where now
\begin{equation}\begin{split} \label{e.blf}
g(\mu,T_e)
=&\frac{D^2 \gamma_1^3 k_B}{4\pi\rho\hbar^5 v^6} 
\left\{2 k_BT_e\ln[2\cosh(\mu/2k_BT_e)]\right\}.
\end{split}\end{equation}
In contrast to MLG, 
if $k_{\rm B}T_{\rm BG,BLG}\ll k_BT_e\ll|\mu|$ then 
$g(\mu,T_e)\propto \mu$,
and if $|\mu|,2(c/v)\sqrt{\gamma_1k_{\rm B}T_{e}}\ll k_BT_e$ then
$g(\mu,T_e)\propto k_BT_e$. 

Finally we mention also the BLG result for the high-temperature 
relaxation rate,
\begin{equation}\begin{split} \label{e.blrateht}
\tau_{k_F}^{-1}=\frac{D^2\gamma_1k_B}{4\rho\hbar^3v^2c^2}T,
\end{split}\end{equation}
which is valid in the same limit as assumed above.


\section{Optical phonons} \label{s.eop}

For graphene, its multilayers, and graphite the phonon spectra have
been studied in detail both experimentally and
theoretically.\cite{mohr07,yan08,michel08,malard09,kuminsky09} In
order to describe the crossover from acoustic phonons to optical
phonons as the dominating phonon type for heat dissipation, we use
some simplified optical-phonon models.  
For the description of the intrinsic optical modes an atomistic
description is needed as a starting point. However, we skip the
details  here (see App.\ \ref{s.opdetails}), 
as similar calculations have been reported 
earlier. Of the intrinsic phonons, we consider explicitly only in-plane LO and TO modes
(collectively LT) at long wavelengths, \emph{i.e.}\ near the $\Gamma$
point.\cite{suzuura02,andoreview05,tsedassarma07,castroneto09,ando09}
(The K-point intrinsic phonons\cite{suzuura08} are expected to have
coupling constants that differ only by a numerical
prefactor.\cite{rana09}) Additionally, we consider coupling to the
``remote'' phonons of a dielectric
substrate,\cite{wang72,fischetti01,fratini08} which is also a relevant
concern for many experiments.\cite{meric08}


\subsection{Coupling constants for simple phonon models} \label{s.optmod}

Let us first consider a model for the long-wavelength in-plane (LT)
optical phonons in MLG, for which the LO and TO branches are nearly
degenerate, with energy $\Omega_{LT}\approx 0.2$ eV.  In the
tight-binding picture of Fig.\ \ref{f.bilayer_struct}, the
strongest coupling to the LT modes comes from the modulation of the
nearest-neighbor (A1-B1) coupling amplitude $\gamma_0$ (see
Ref.\ \onlinecite{papanote}).  The coupling constant is of the
form\cite{tsedassarma09}
\begin{equation} \label{e.mlltccx}
w_{\vec{k}\vec{p},\vec{q}}^{\alpha\beta,LT}
=
\frac{9(\gamma_0')^2\hbar^2}{2M\Omega_{LT}}
\frac{1}{2}(1-s_\alpha s_\beta 
\cos(\phi_{\vec{k}} + \phi_{\vec{p}} -2\phi_{\hat{\vec{a}}})),
\end{equation}
where $M=A\rho_{1}$, with $\rho_{1}\approx 7.6\cdot 10^{-7}$ kg/m$^2$ 
the mass density of MLG, and $\gamma_0'\approx$ 40 eV/nm is the
derivative of $\gamma_0$ with respect to the nearest-neighbor bond length.\cite{suzuura02} 
The vector
$\hat{\vec{a}}(\vec{q})=\hat{\vec{q}}$ for the LO and
$\hat{\vec{a}}(\vec{q})=\hat{\vec{z}}\times\hat{\vec{q}}$
for the TO branch, $\hat{\vec{z}}$ being normal to the plane. 

In BLG there are four nearly degenerate LT branches, with
$\Omega_{LT}\approx 0.2$ eV, since for both LO and TO type modes the
atoms in layers 1 and 2 can move either in phase or in opposite
phases. \cite{ando09} With perturbation theory to first order in
$\hbar v k/\gamma_1$ we find in this case
\begin{equation}\begin{split} \label{e.blltccx}
w_{\vec{k}\vec{p},\vec{q}}^{\alpha\beta,LT}
= &
\frac{9(\gamma_0')^2\hbar^2}{2M\Omega_{LT}}
\frac{1}{2}\frac{(\hbar v)^2}{\gamma_1^2}
\bigg\{
k^2 + p^2 \\
& + 2kp
\left[
\pm\cos(\phi_{\vec{k}\vec{p}})
- s_{\alpha}s_{\beta}\cos(\phi_{\vec{k}\hat{\vec{a}}} + \phi_{\vec{p}\hat{\vec{a}}})
\right] \\
&\mp s_{\alpha}s_{\beta}
\left[
k^2\cos(2\phi_{\vec{p}\hat{\vec{a}}})
+
p^2\cos(2\phi_{\vec{k}\hat{\vec{a}}})
\right]
\bigg\}
\end{split}\end{equation} 
where $M=A\rho_{2}$, $\rho_{2}=2\rho_{1}$,  
and $\phi_{\vec{k}\vec{p}}=\phi_{\vec{k}}-\phi_{\vec{p}}$. 
The upper and lower signs are for the in-phase and opposite-phase modes, respectively.
For BLG there are also the ZO modes with $\Omega_{ZO}\approx 0.1$ eV which
could couple linearly to electrons via the modulation of the A1-B2 bond and thus 
$\gamma_1$, but  we estimate them to be relatively unimportant for the energy
relaxation (see Ref.\ \onlinecite{zonote}).

Finally, in addition to these intrinsic phonons, we consider
possibility of coupling to remote surface phonons of a dielectric
substrate.\cite{fischetti01,fratini08,meric08} The coupling is due to
the electric polarization associated with the phonons, which modulates
the scalar potential on the graphene.  For simplicity we again
disregard the possibility of intervalley scattering.\cite{fratini08}
The coupling constant is given by\cite{wang72,fratini08}
\begin{equation} \label{e.remcc}
w_{\vec{k}\vec{p},\vec{q}}^{\alpha\beta,\textrm{rem}} = 
\beta_{\rm rem}
\frac{e^2\hbar\Omega_{\textrm{rem}}}{2\varepsilon_0 A}
\frac{1}{q} e^{-2qz}
F_{\alpha\beta}(\theta),
\end{equation}
where 
$\Omega_{\rm rem}$ is the energy of the relevant surface mode, 
$z\geq0$ is the effective distance between the graphene and the substrate,
$F_{\alpha\beta}(\theta)$ is as in Eq.\ (\ref{e.gaccoupling}),
and $\varepsilon_{0}$ is the permittivity of vacuum in SI units.
If there is only a single relevant surface mode, 
$\beta_{\rm rem}=(\varepsilon_s-\varepsilon_\infty)/[(\varepsilon_s+1)(\varepsilon_\infty+1)]$, where 
$\varepsilon_s$ and $\varepsilon_\infty$ are the static and high-frequency dielectric 
constants of the insulator.\cite{wang72} If there are several surface modes,
Eq.\ (\ref{e.remcc}) should be additionally weighted by the 
corresponding relative oscillator strengths.\cite{fratini08}


\subsection{Heat current between electrons and optical phonons} \label{s.optres}

When the momentum conservation is imposed, $q=\sqrt{k^2+p^2-2kp\cos\theta}$, where
$\theta=\phi_{{\vec{k}}}-\phi_{{\vec{p}}}$. All the coupling constants may then be expressed 
in terms of only $k$, $p$, and $\cos\theta$.
For an electron dispersion of the form $\epsilon_k^\alpha=s_\alpha\epsilon_k$, $\epsilon_k>0$, the 
heat current between electrons and dispersionless optical phonons may be written 
\begin{equation}
Q_{e-op}^{(\gamma)} = Aq_{e-op}^{(\gamma)}(\mu,T_e)[n_{e}(\Omega_\gamma)-n_{ph}(\Omega_\gamma)],
\end{equation}
where (including degeneracy of $g_e=4$)
\begin{equation}\begin{split}  \label{e.geop}
q_{e-op}^{(\gamma)} =&   
\frac{A}{4\hbar}\sum_\gamma\Omega_\gamma^2
\int_{-\infty}^{\infty} \nu(\Omega_\gamma x)\nu(\Omega_\gamma (x-1)) \\
&\times\int_{-\pi}^{\pi}d\theta 
w^{\alpha\beta,\gamma}(k(\Omega_\gamma|x|),k(\Omega_\gamma|x-1|),\cos\theta) \\
&\times[f(\Omega_\gamma(x-1))-f(\Omega_\gamma x)]dx.
\end{split}\end{equation}
Here 
$\alpha\beta$ are chosen such that 
$s_\alpha=\mathrm{sign}(x)$ and $s_\beta=\mathrm{sign}(x-1)$,
$\nu(\epsilon)$ is the density of electronic states, and $k(\epsilon)$ the inverse of $\epsilon_k$.
The interval $0<x<1$ corresponds to interband scattering.

In the case of the LT modes further simplification is achieved by
summing over the degenerate LT modes $\gamma$. The angle-dependent
terms of the coupling constants (\ref{e.mlltccx}) or (\ref{e.blltccx})
then cancel and the angle integral in Eq.\ (\ref{e.geop}) becomes
trivial.  Thus for MLG\cite{bistritzer09}
\begin{equation}
q_{e-op}^{(LT)}(\mu,T_e) = \frac{9\Omega^3 (\gamma_0')^2\hbar}
{\pi (\hbar v)^4\rho_{1}}\mathcal{F}(\mu,T_e),
\end{equation}
where the factor $g_e=4$ is included, $\Omega=\Omega_{LT}$, and 
\begin{equation} \label{e.ffu}
\mathcal{F}(\mu,T_e) = \int_{-\infty}^{\infty}|x(1-x)|
[f(\Omega(x-1)) - f(\Omega x) ]dx.
\end{equation}
For $\mu=0$ and $\Omega/k_BT_e\gg 1$ it has the value
$\mathcal{F}=1/6$.
Similarly for BLG
\begin{equation} \label{e.blqlto}
q_{e-op}^{(LT)}(\mu,T_e)
=\frac{18\Omega^3 (\gamma_0')^2\hbar}{\pi (\hbar v)^4\rho_{2}}
\frac{\gamma_1}{\Omega}\mathcal{G}(\mu,T_e),
\end{equation}
where 
\begin{equation} \label{e.gfu}
\mathcal{G}(\mu,T_e) = \int_{-\infty}^{\infty}\frac{1}{4}(|x|+|x-1|)
[f(\Omega(x-1)) - f(\Omega x) ]dx.
\end{equation}
For $\mu=0$ and $\Omega/k_BT_e\gg 1$ this function has the value
$\mathcal{G}=1/4$. We note that due to the assumed parabolic
dispersion, this result is a good approximation only close to
$|\mu|=0$.  For the case of the remote phonons, the angle integral in
Eq.\ (\ref{e.geop}) is quite complicated, but in the limit
$z\rightarrow 0$ it may be carried out analytically for both MLG and
BLG. However, the expressions are complicated and we skip them here.

\begin{figure}[!tb] 
\includegraphics[width=0.9\linewidth,clip=]{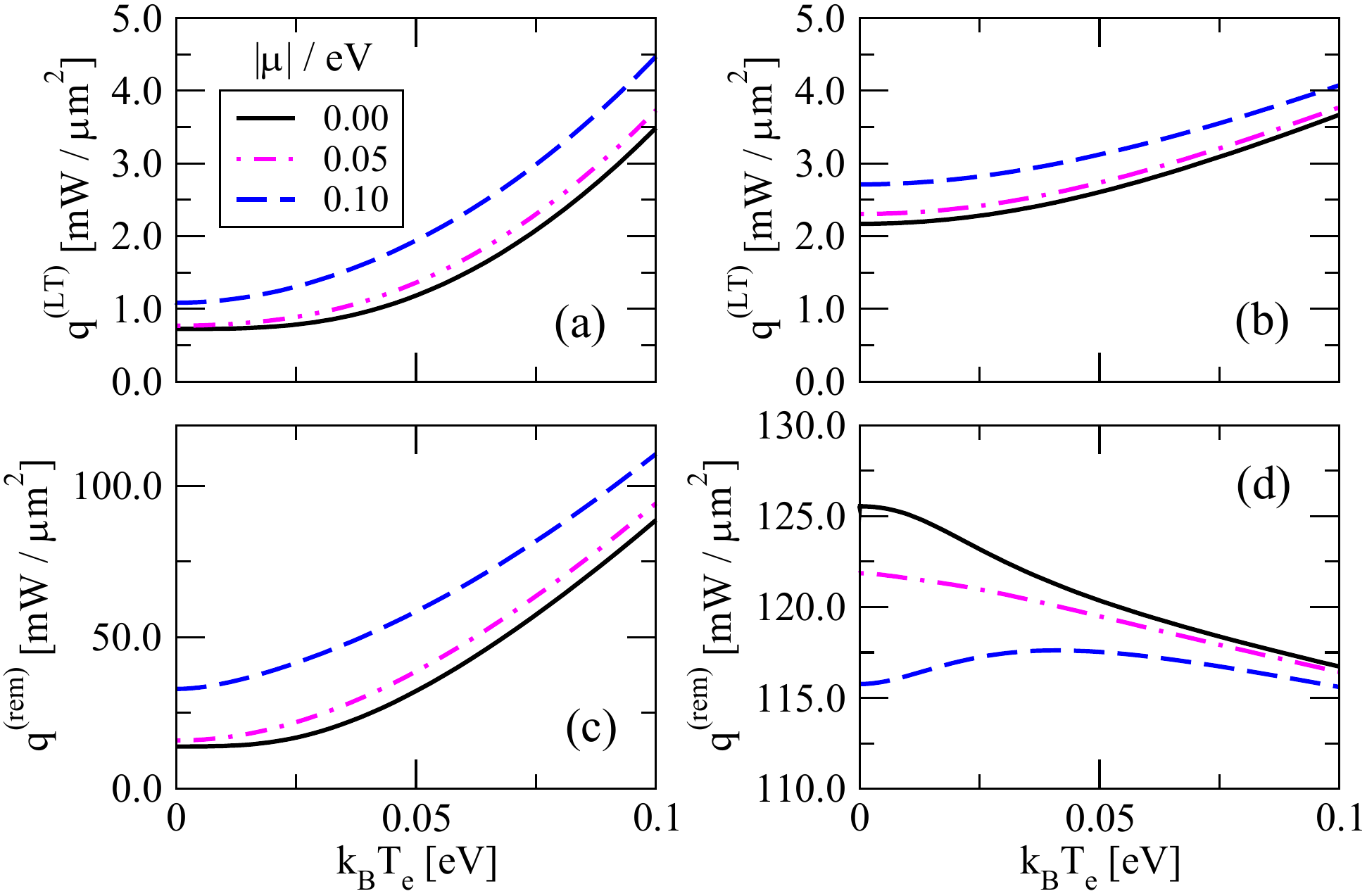}
\caption{(Color online) Prefactors $q_{e-op}^{(LT)}$ and
  $q_{e-op}^{({\rm rem})}$ of the electron-optical phonon heat
  currents for different doping levels.  (a) and (c) are for MLG, and
  (b) and (d) for BLG. Upper panels are for the intrinsic LT phonons
  and the lower ones for the remote phonons, with $z=0$.}
\label{f.qfactop}
\end{figure}

The numerical results for the prefactors $q_{e-op}^{(\gamma)}$ are
shown in Fig.\ \ref{f.qfactop}.  For the remote phonons we use $z=0$
and the values for 6H-SiC (see Ref.\ \onlinecite{nienhaus95}), where
$\varepsilon_s=9.72$, $\varepsilon_\infty=6.52$,
$\Omega_{\textrm{rem}}= 116$ meV, and hence $\beta_{\rm rem}=0.040$.
It is seen from Fig.\ \ref{f.qfactop} that the prefactors for the
remote phonons are considerably higher than those of the intrinsic
phonons.  This fact combined with $\Omega_{\rm rem}<\Omega_{LT}$ means
that for heat dissipation the remote phonons may in practice always be
the more important ones.  The limit $z=0$ overestimates the coupling
constant somewhat, but $q_{e-op}^{(\rm rem)}$ decays with distance at most
proportionally to $e^{-2q_{\rm max}z}$, where for example for MLG at
$k_{\rm B}T_e\ll\Omega_{\rm rem}$ we find 
$q_{\rm max}=(\Omega_{\rm rem}+2|\mu|)/(\hbar v)$ 
and thus $q_{\rm max}^{-1}\approx 6.2$ nm
when $\mu=0$.  The use of SiO$_2$ would reduce
$\varepsilon_{s}-\varepsilon_{\infty}$, but then one of the two modes
with appreciable oscillator strengths also has a lower
energy.\cite{fischetti01} The corresponding crossover temperatures
between $Q_{e-ac}$ and $Q_{e-op}$ are discussed below.

\section{Discussion and conclusions} \label{s.discussion}

Let us briefly discuss the implications of our results. A common
experimental situation where the heating of electrons occurs is that
of a two-probe measurement with a finite bias voltage $V$. The Joule
heat created at some point of the system is dissipated from the
electrons in basically two ways, either by a transfer to phonons or by
diffusion of the electrons away from the heated region ({\it e.g.},
into the electrodes). The static situation is described by a heat
balance equation $\nabla\cdot\vec{j}_Q+P_{\rm e-ac}=P_J$, where 
$\vec{j}_Q$ is the heat current density, while 
$P_{\rm e-ac}$ and $P_{J}$ are the local electron-acoustic phonon power and
Joule power per area, respectively.  We assume for simplicity that the
Wiedemann-Franz law applies (see Ref.\ \onlinecite{trushin07})
so that $\vec{j}_Q=-\kappa\nabla T_e$, where $\kappa=\mathcal{L}\sigma T_e$ 
is the heat conductivity, $\sigma$ the electrical conductivity and
$\mathcal{L}=(\pi^2/3)(k_B/e)^2$ is the Lorenz number. To find some
order-of-magnitude estimates, we assume a quasi-one-dimensional
situation with a sample of length $L$, and consider the simpler, 
discretized equation $P_{\rm diff} + P_{\rm e-ac} = P_J$ with the diffusion power
$P_{\rm diff}=(4\mathcal{L}\sigma/L^2)(T^2-T_0^2)$ and Joule power
$P_{J}=\sigma V^2/L^2$.  We also concentrate on the low-temperature
regime, $k_BT\ll k_BT_{\rm BG,MLG}=2(c/v)|\mu|$ for MLG and $k_BT\ll
k_{B}T_{\rm BG,BLG}=2(c/v)\sqrt{\gamma_1|\mu|}$ for BLG, so that
$P_{\rm e-ac}$ is of the form $P_{\rm e-ac}=\Sigma(T^4-T_0^4)$.  Here
$T$ is the electron temperature in the middle of the
graphene sample and we assume the acoustic phonons and the electrons
in the leads to remain at the bath temperature $T_0$.  

Clearly, at low enough $T_{0}$ and bias $V$ the diffusion power dominates over
the phonon power ($P_{\rm e-ac}<P_{\rm diff}$). However, with increasing $T_0$ or $V$ 
there is a crossover to an electron-phonon dominated regime
($P_{\rm e-ac}>P_{\rm diff}$).  If $T_0>T_{0x}=2\mathcal{L}\sigma/(\Sigma L^2)$ 
then $P_{\rm e-ac}$ dominates also in the linear-response regime (arbitrarily small $V$ and 
$\Delta T=T-T_0$).  For $T_0<T_{0x}$ there is a finite crossover voltage 
$V_{\rm cr}=\sqrt{8\mathcal{L}(T_{\rm cr}^2-T_0^2)}$, 
with the corresponding
temperature $T_{\rm cr}=\sqrt{4\mathcal{L}\sigma/(\Sigma L^2)-T_0^2}$.

When the phonon power dominates, $P_{\rm e-ac}\approx P_{J}$.  Since
$P_{J}$ can be deduced from the current-voltage characteristics,
measurement of the electron temperature in the presence of heating can
also act as an indirect measurement of the electron-phonon coupling
constant. To see if this regime can be reached before the
low-temperature approximation for $P_{\rm e-ac}$ breaks down, we
estimate the crossover temperatures.  Using $D=30$ eV we find for MLG
\begin{equation} \label{e.mlx}
T_{0x}=5~\mathrm{K}\times\frac{(\sigma/\sigma_{0})^{1/2}}
{(L/1~\mu\mathrm{m})(|\mu|/0.3~\textrm{eV})^{1/2}}
\end{equation}
and for BLG
\begin{equation} \label{e.blx}
T_{0x}=11~\mathrm{K}\times
\frac{(\sigma/\sigma_{0})^{1/2}(|\mu|/0.3~\textrm{eV})^{1/4}}
{(L/1~\mu\mathrm{m})},
\end{equation}
where $\sigma_0=4e^2/h$.
These are only valid if $T_{0x}\ll T_{\rm BG, MLG/BLG}$.
We find $T_{\rm BG,MLG}=140~\mathrm{K}\times (|\mu|/0.3~\textrm{eV})$
for MLG
and $T_{\rm BG,BLG}=160~\mathrm{K}\times (|\mu|/0.3~\textrm{eV})^{1/2}$ for BLG.
Since the electron-phonon interaction becomes more important for 
increasing $L$, it seems possible to meet these criteria with 
long enough samples.

The estimate could be improved by taking into account that $\sigma$
also grows with $\mu$ differently for MLG and BLG, and it may also
depend on temperature.\cite{bolotin08,muller09} However, these details
depend on the types of scattering. In practice, the dominant form of
scattering in graphene on a substrate is from
impurities.\cite{adam08,dassarma09} Experiments indicate that $\sigma$
grows roughly linearly with charge density,\cite{morozov08} which
yields $\sigma\sim\mu^2$ for MLG and $\sigma\sim\mu$ for BLG.  We also
note that the diffusion power can be reduced or even eliminated by the
use of superconducting leads.

\begin{figure}[!tb] 
\includegraphics[width=0.9\linewidth,clip=]{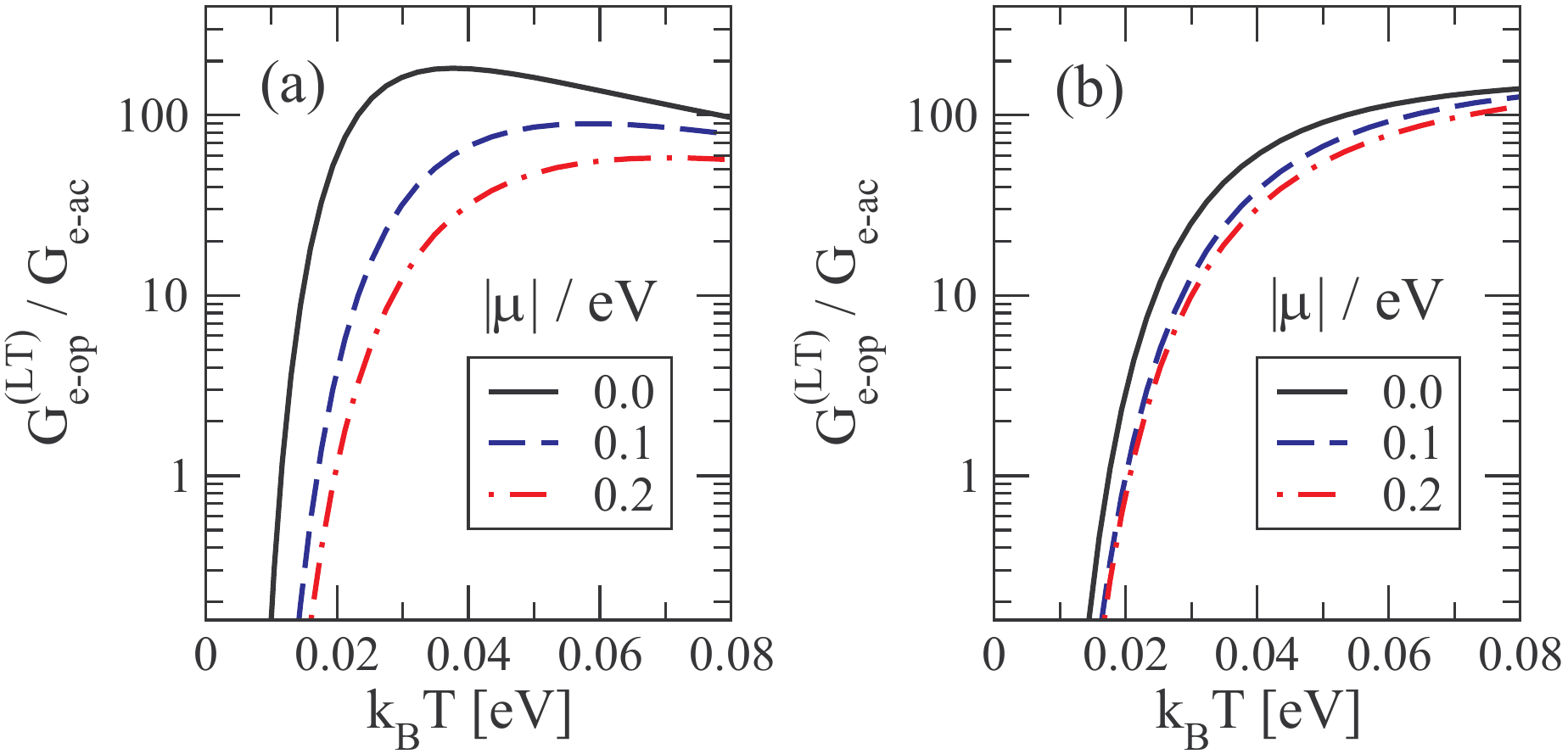}
\caption{(Color online) The ratio of (LT) optical phonon thermal
  conductance $G_{e-op}^{(LT)}(\mu,T)$ and that of acoustic phonons
  $G_{e-ac}(\mu,T)$ for MLG (a) and BLG (b). The parameters are $D=30$
  eV and $|\gamma_0'|=40$ eV/nm (Ref.\ \onlinecite{suzuura02}).  In
  both cases the crossover point $G_{e-op}^{(LT)}/G_{e-ac}=1$ moves to
  higher temperature with increasing $|\mu|$.}
\label{f.ratioplots}
\end{figure}

At high enough temperature there is another crossover, where optical
phonons begin to dominate the electron-phonon heat
transfer.\cite{bistritzer09,tsedassarma09} Although the conductance
$G_{e-op}$ is exponentially suppressed at low temperature, $G_{e-ac}$
is also small and for MLG quite strongly temperature-dependent, with
$G_{e-ac}\sim T^4$ at $\mu=0$. This can lead to surprisingly low
crossover temperatures. Here we consider explicitly only the intrinsic
LT optical phonon modes.  Figure \ref{f.ratioplots} shows the ratios
$G_{e-op}^{(LT)}/G_{e-ac}$ of optical and acoustic phonon thermal
conductances for MLG and BLG. As found
previously,\cite{tsedassarma09,bistritzer09} optical phonons can
become dominant already well below room temperature ($150$ K for MLG
and $200$ K for BLG), and the crossover moves to higher temperatures
with increasing carrier density.  The non-monotonous
behavior\cite{bistritzer09} of $G_{e-op}^{(LT)}/G_{e-ac}$ in the case
of MLG is not present in BLG.  As found above, if the graphene is on a
dielectric substrate, the dominant optical modes are most likely the
surface modes of the dielectric.\cite{fratini08,meric08} By using the
parameter values quoted in Sec.\ \ref{s.optres} for SiC and studying
$G_{e-op}^{(\rm rem)}/G_{e-ac}$ as in Fig.\ \ref{f.ratioplots}, we find
crossover temperatures $60$ K for MLG and $80$ K for BLG at $\mu=0$.
For SiO$_2$, using the parameters quoted in
Ref.\ \onlinecite{fratini08}, the results are as low as $30$ K and
$50$ K, respectively.

We note that the above results depend on the poorly-known parameters
$D$ and $\gamma_0'$ in their second powers, while $c$ and $v$ also
appear with high powers.  This makes quantitative predictions difficult,
and increases the need for an experimental determination of the
coupling constants.  Finally it should also be noted that if the bias
voltage exceeds the value $\sim \Omega_{LT}/e=0.2$ V, scattering from
optical phonons becomes very important also at low bath
temperature.\cite{meric08,barreiro09,chauhan09} In this case the
optical phonons can have a highly non-equilibrium
distribution.\cite{auer06,lazzeri06,song08}

To conclude, we have calculated the power transfer between the
electron and phonon systems in monolayer and bilayer graphene,
assuming the existence of quasiequilibrium.  In particular we have
studied the coupling to longitudinal acoustic phonons and different
types of optical phonons. For the former we have calculated the power
numerically and derived analytic expressions in low- and
high-temperature limits.  The power transfer to acoustic phonons
dominates the diffusion power above a certain crossover temperature,
estimated in Eqs.\ (\ref{e.mlx}) and (\ref{e.blx}).  At even higher
temperatures, there is another crossover where optical phonons begin
to dominate, and we have estimated also these crossover temperatures
numerically (Fig.\ \ref{f.ratioplots}).  We find that for graphene on
the substrate the most relevant optical phonons are likely to be the
surface optical modes of the substrate.

\acknowledgments

We acknowledge the useful discussions with Aurelien Fay, Pertti
Hakonen, and Francesco Giazotto. This work was supported by the 
Academy of Finland, the European Research Council Starting Grant 
(Grant No.\ 240362-Heattronics), and the NANOSYSTEMS/Nokia contract 
with the Nokia Research Center.


\appendix


\section{Analytic solutions for the conservation laws in scattering of electrons
from acoustic phonons} \label{s.anasol}

\begin{figure}[!tb] 
\includegraphics[width=0.5\linewidth,clip=]{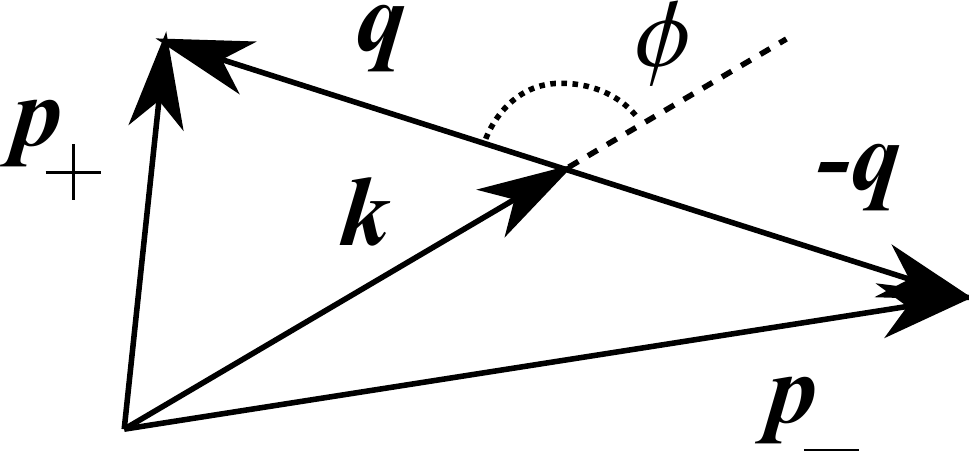}
\caption{Momentum conservation in absorption ($+$) or emission ($-$)
  of a phonon. $\vec{k}$ is the incoming electron wave vector,
  $\vec{p_{\pm}}$ the outgoing one, and $\vec{q}$ that of the
  associated phonon.  }
\label{f.arrows}
\end{figure}

\begin{figure}[!tb] 
\includegraphics[width=0.8\linewidth,clip=]{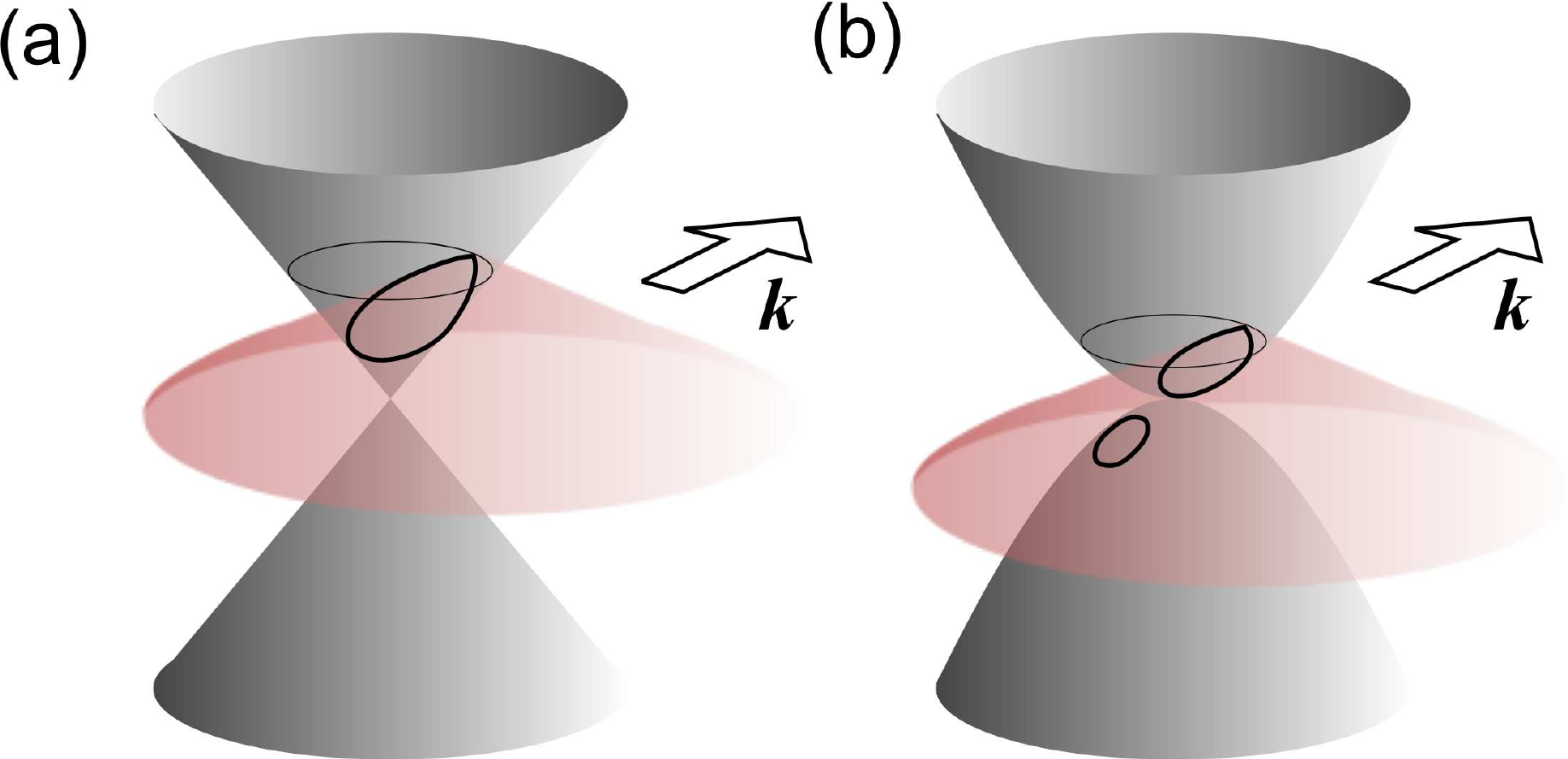}
\caption{(Color online) Schematic illustration ($c/v$ not to scale) of
  the allowed final states in the momentum-energy space for
  electron-acoustic phonon interaction in the case of MLG (a) and BLG
  (b).  These are given by the curves along which the two
  momentum-energy surfaces intersect (thick solid lines).  The curves
  corresponding to elastic scattering are also shown (thin solid
  line).  The initial state (with $\vec{k}$ in the direction of the
  arrow) is in the conduction band and only the case of emission is
  shown.  In the case of MLG, no interband scattering is possible.  }
\label{f.shaded}
\end{figure}

Here we detail the solutions for the possible
final states allowed by momentum and energy conservation laws
in electron-acoustic phonon scattering.
If $\vec{k}$ and $\vec{p}$ are the initial and final wave vector of
the electron, respectively, $\vec{q}$ the wave vector of the associated phonon, and
$k=|\vec{k}|$, $p=|\vec{p}|$, and $q=|\vec{q}|$,  then
the conservation laws are
\begin{equation}\begin{split} \label{e.claws}
\vec{p} &= \vec{k} + s\vec{q}, \\
\epsilon_p^\beta &= \epsilon_k^\alpha + s\hbar c q.
\end{split}\end{equation}
Here $s=+1$ for absorption and $s=-1$ for emission,
and $\alpha = {\rm c,v}$ is the band index.  
The final wave vectors $\vec{p}_s=\vec{k}+s\vec{q}$
are illustrated in Fig.\ \ref{f.arrows}.
Figure \ref{f.shaded} illustrates schematically the 
allowed states in momentum-energy space.

\begin{figure}[!tb] 
\includegraphics[width=0.9\linewidth,clip=]{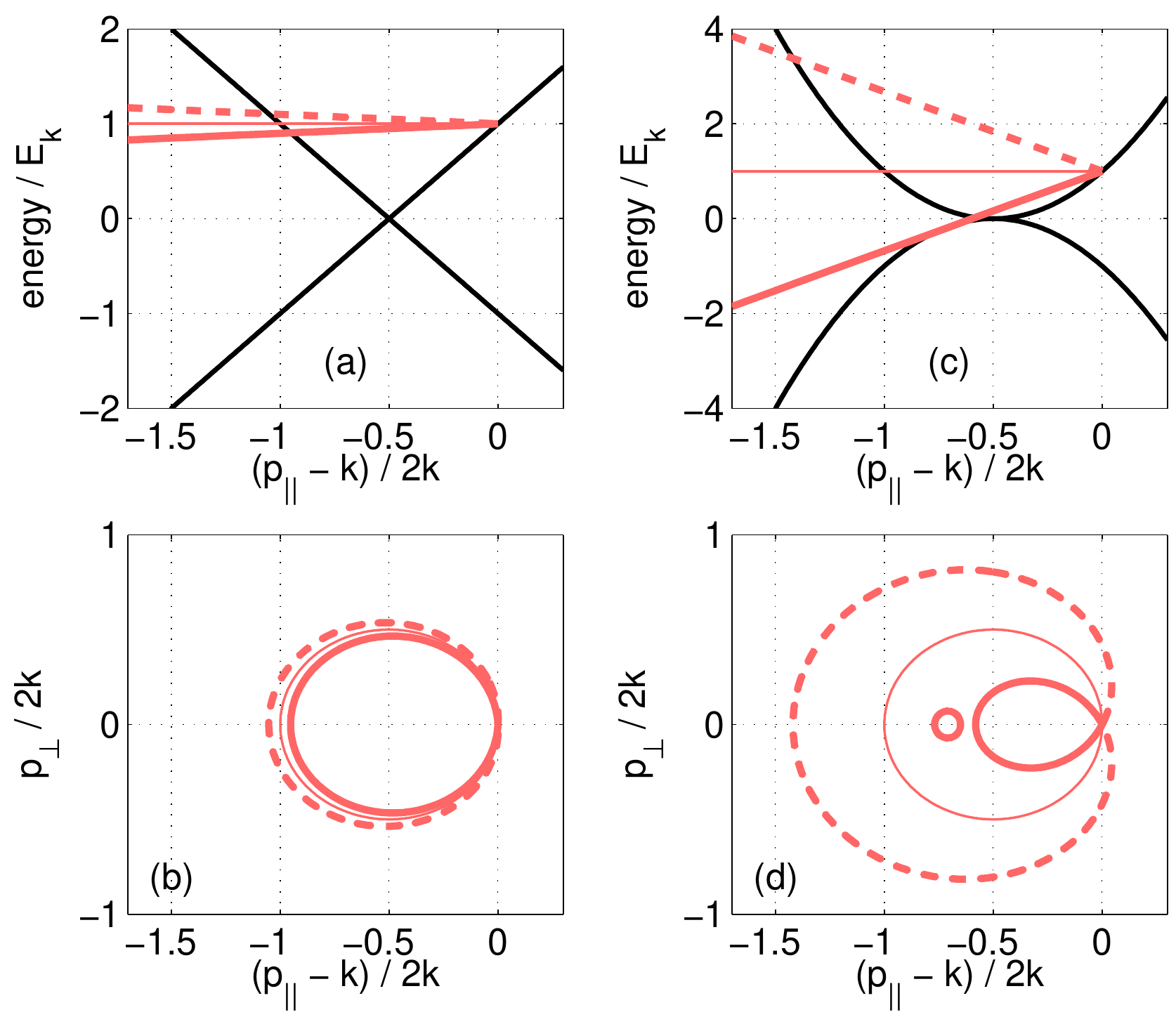}
\caption{(Color online) Allowed final states in the $\vec{p}$ plane
  for electron-acoustic phonon interaction in the case of MLG (a,b)
  and BLG (c,d) graphene.  The upper diagrams (a,c) show the
  intersections of Fig.\ \ref{f.shaded} in $(p_\parallel,E)$ plane,
  where we define $p_\parallel$, $p_\perp$ as the components of
  $\vec{p}$ parallel and perpendicular to $\vec{k}$.  The lower
  diagrams (b,d) show the curves in $\vec{p}$ plane.  The thin lines
  correspond to elastic scattering, thick solid lines to emission, and
  thick dashed lines to absorption.  Interband scattering is possible
  only for BLG, and here only for emission (initial state in the
  conduction band).  For MLG we use the unphysically large $c/v=0.05$
  for clarity of the figure, and for BLG
  $x_c(k)=x_c(k_{c1})+0.005=\sqrt{2}-1+0.005$.  Energies are in units
  of $E_k=\hbar v k$.  }
\label{f.polplot}
\end{figure}

We represent the final states in polar coordinates by 
writing $\vec{p}_s=\vec{k}+s(q\cos\phi \hat{\vec{p}}_\parallel + q\sin\phi \hat{\vec{p}}_\perp)$, where
$\hat{\vec{p}}_\parallel$ is the unit vector parallel to $\vec{k}$ and
$\hat{\vec{p}}_\perp$ the one perpendicular to it, and $\phi$ is the
angle between $\vec{q}$ and $\vec{k}$ (see Fig.\ \ref{f.arrows}).  
The curves are then described parametrically by the functions 
$q=q(k,\cos\phi)$. We give the solutions first for
MLG and then for BLG.
The results are illustrated in Fig.\ \ref{f.polplot}.

\subsection{Monolayer graphene}

For MLG $\epsilon_k^\alpha = s_\alpha \hbar v k$,
with $s_\alpha=\pm 1$. Eliminating $\vec{p}$ in Eq.\ (\ref{e.claws}),
we find
\begin{equation}\begin{split}
s_\beta\hbar v p_s(k,q,x) &= 
s_\alpha \hbar v k + s\hbar c q,
\end{split}\end{equation}
where $p_s(k,q,x)=|\vec{p}_s|=\sqrt{k^2 + q^2 + 2kqsx}$,
with $\vec{p}_s=\vec{k}+s\vec{q}$ ($s=\pm 1$),
$x=\cos\phi=\vec{q}\cdot\vec{k}/(kq)$,
and $s=\pm 1$ for absorption or emission, respectively.
This yields the parametric representation for the curves
\begin{equation}\begin{split}
\frac{q}{2k} = 
\begin{cases}
\frac{-sx+ss_\alpha(c/v)}{1-(c/v)^2}, & sx < ss_\alpha(c/v),~\beta=\alpha \\
0, & sx > ss_\alpha(c/v),~\beta=\alpha \\
\textrm{no real solution}, & \beta\neq\alpha. \\
\end{cases}
\end{split}\end{equation}
There are thus no finite-$q$ solutions if $sx>c/v$,
one for 
$-c/v < sx < c/v$ and two if $sx < -c/v$, so that 
$\lim_{q\rightarrow 0}|x|=c/v$.
The maximal values are (set $sx=-1$)
\begin{equation}\begin{split}
\frac{q_{\rm max}^{\alpha\beta}}{2k} = 
\begin{cases}
\frac{1}{1-ss_\alpha(c/v)}, & \beta=\alpha \\
\textrm{no real solution}, & \beta\neq\alpha, \\
\end{cases}
\end{split}\end{equation}
while the minimal value is always $q_{\rm min}^{\alpha\alpha}=0$.
In the special case $c/v=0$ the scattering is elastic and $q=-2ksx$,
which has the maximal value $2k$ corresponding to
backscattering.
If $s_\alpha=+1$, then $q_{\rm max}^{\alpha\alpha}<2k$ 
for emission ($s=-1$) and 
$q_{\rm max}^{\alpha\alpha}>2k$ for absorption ($s=+ 1$), and vice
versa for $s_\alpha=-1$.


\subsection{Bilayer graphene}

For a BLG $\epsilon_k^\alpha = s_\alpha (\hbar v k)^2/\gamma_1$,
with $s_\alpha=\pm 1$. Now
\begin{equation}\begin{split}
s_\beta(\hbar v p_s(k,q,x))^2/\gamma_1
&= s_\alpha (\hbar v k)^2/\gamma_1 + s\hbar c q,
\end{split}\end{equation}
where again $p_s(k,q,x)=\sqrt{k^2 + q^2 + 2kqsx}$,
with
$x=\cos\phi=\vec{k}\cdot\vec{q}/(kq)$.
Now we find the following solutions. For $\beta=\alpha$
\begin{equation}\begin{split}
\frac{q}{2k} = 
\begin{cases}
-sx+ss_\alpha x_c(k), & sx < ss_\alpha x_c(k),~k>k_{c2} \\
0, & sx > ss_\alpha x_c(k),~k>k_{c2} \\
-sx+x_c(k), & 0<k\leq k_{c2},~ss_\alpha = +1 \\
0, & 0<k\leq k_{c2},~ss_\alpha = -1, \\
\end{cases}
\end{split}\end{equation}
where $x_c(k)=(1/2)(c/v)(\gamma_1/\hbar v k)$
and $k_{c2}=(1/2)(c/v)(\gamma_1/\hbar v)$, so that
$x_c(k_{c2})=1$. 
For $k>k_{c2}$ there are thus no finite-$q$ solutions if $sx>x_c(k)$,
one for $x_c(k) < sx < x_c(k)$ and two if $sx < -x_c(k)$,
such that $\lim_{q\rightarrow 0}|x|=x_c(k)$.
No solutions exist for $k<k_{c1}$. 
The extremal values of $q$ are
\begin{equation}\begin{split}
\frac{q_{\rm max,min}^{\beta=\alpha}}{2k} = 
\begin{cases}
1+ss_\alpha x_c(k),~0 &
k>k_{c2} \\
\pm 1 + x_c(k), & 0<k\leq k_{c2},~ss_\alpha = +1 \\
0, & 0<k\leq k_{c2},~ss_\alpha = -1, \\
\end{cases}
\end{split}\end{equation}
where the maximum (minimum) follows by setting $sx=-1$ ($sx=+1$).
Again, $q_{max}^{\beta=\alpha}\gtrless 2k$ for absorption/emission, 
if $s_\alpha=+1$ and vice versa for $s_\alpha=-1$.

For $\beta\neq\alpha$ we have
\begin{equation}\begin{split} \label{e.intersol}
\frac{q}{2k} = 
\frac{1}{2}\left[-sx+x_c(k)\pm\sqrt{(-sx+x_c(k))^2-2}\right], 
\end{split}\end{equation}
if $|sx-x_c(k)|\geq\sqrt{2}$, $0<k\leq k_{c1}$, $ss_\alpha=-1$ 
where $k_{c1}=k_{c2}/(\sqrt{2}-1)>k_{c2}$.
For $k>k_{c1}$ or $ss_\alpha=+1$ no real solution exists.
In this case there are only solutions for 
either emission ($s=-1$) or absorption ($s=+1$), 
depending on $s_\alpha=-s_\beta$
such that $s s_\beta = -ss_\alpha = +1$.
The upper and lower signs in Eq.\ (\ref{e.intersol}) describe the
two branches of the interband solution,
represented by the closed circular curve 
in Fig.\ \ref{f.polplot}.
By setting $sx=-1$, we find that the 
extremal values for $q$ are now
\begin{equation}\begin{split}
\frac{q_{\rm max,min}^{\beta\neq\alpha}}{2k} = 
\frac{1}{2}\left[1+x_c(k)\pm\sqrt{(1+x_c(k))^2-2}\right], 
\end{split}\end{equation}
if $0<k\leq k_{c1}$ and $ss_\alpha=-1$. 
We also note that $x_c(k_{c1})=\sqrt{2}-1\approx 0.41421$ and thus
$q_{c1,{\rm max}}^{\beta\neq\alpha}/2k_{c1}=q_{c1,{\rm min}}^{\beta\neq\alpha}/2k_{c1}=1/\sqrt{2}\approx 0.70711$.
The corresponding value for the intraband solution
($ss_\beta=ss_\alpha=-1$) 
is $q_{{\rm max},c1}^{\beta=\alpha}/2k_{c1}=2 -\sqrt{2}\approx 0.58579$.


\section{Atomistic description of electron-phonon coupling in graphene} \label{s.opdetails}

Here we consider the tight-binding description of the coupling of
electrons to the intrinsic phonons in graphene, and explain in more
detail our simple models for the long-wavelength optical phonons in
MLG or BLG. We only consider BLG in detail.

\begin{figure}[!tb] 
\includegraphics[width=0.9\linewidth,clip=]{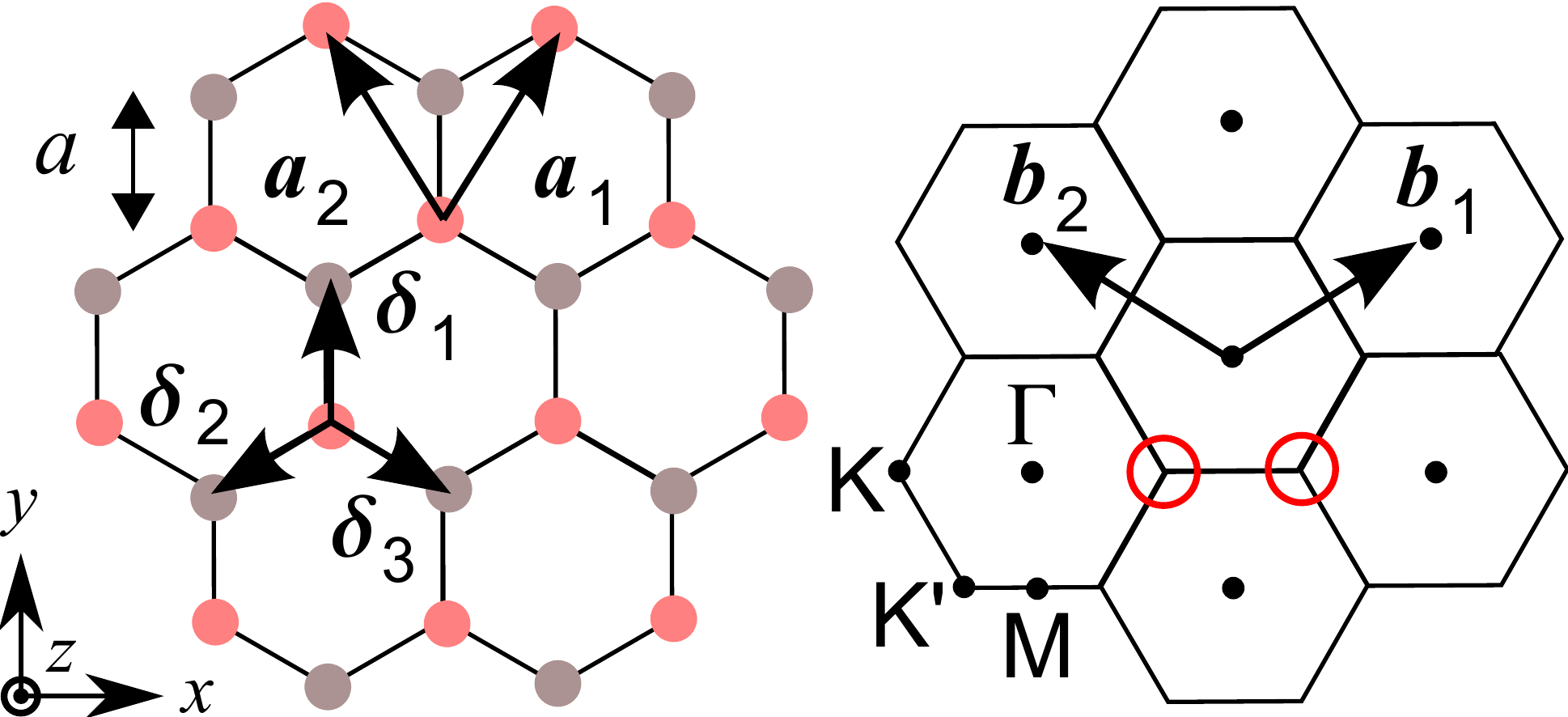}
\caption{(Color online) Direct and reciprocal lattices for graphene,
  showing the primitive vectors $\vec{a}_{1,2}$, $\vec{b}_{1,2}$ of
  the Bravais lattices and the nearest-neighbor vectors
  $\vec{\delta}_{1,2,3}$.  For the reciprocal space the high-symmetry
  points $\Gamma$, K, K', and M are shown, and the hexagons around the
  $\Gamma$ points are the first Brillouin zones.  The two inequivalent
  parts of the Fermi surface are sketched with circles.  }
\label{f.lattices}
\end{figure}

For the unperturbed system (without electron-phonon coupling) 
the tight-binding Hamiltonian of BLG is of the form
\begin{equation}\begin{split} \label{e.bilayerham1}
\hat{H}_e=\sum_{\vec{k}}\Psi_{\vec{k}}^\dagger H(\vec{k}) \Psi_{\vec{k}},
\end{split}\end{equation}
where 
$\Psi_{\vec{k}}^\dagger=(a_{1\vec{k}}^\dagger,
b_{1\vec{k}}^\dagger, b_{2\vec{k}}^\dagger, a_{2\vec{k}}^\dagger)$
consists of the electron creation operators at different sites
and
\begin{equation}\begin{split} \label{e.ham44}
H(\vec{k}) = 
\hbar v k
\left(
\begin{matrix}
0                  & \pm e^{\pm i\phi_{\vec{k}}}  &\gamma_1/\hbar v k       &0 \\
\pm e^{\mp i\phi_{\vec{k}}}  &0               &0               &0 \\
\gamma_1/\hbar v k &0               &0               & \pm e^{\mp i\phi_{\vec{k}}}\\
0                  &0               & \pm e^{\pm i\phi_{\vec{k}}} &0
\end{matrix}
\right),
\end{split}\end{equation}
with $\phi_{\vec{k}}=\arctan(k_y/k_x)$
and $\hbar v=3a\gamma_0/2$.
The upper and lower signs are for the $\vec{K}$
and $\vec{K}'$ points, respectively (see Fig.\ \ref{f.lattices}).
The zero of energy is thus set at the charge neutrality point.
The matrix $H(\vec{k})$ has four eigenvalues (energy bands) and eigenvectors,
which we denote $\epsilon_{\alpha}(\vec{k})$
and $\Phi_{\alpha}(\vec{k})$, $\alpha=1,2,3,4$.
The Hamiltonian (\ref{e.bilayerham1}) and other similar operators 
may be written in the eigenbasis by expanding
$\Psi_{\vec{k}}=\sum_\alpha\Phi_{\alpha}(\vec{k})c_{\vec{k}\alpha}$
where $c_{\vec{k}\alpha}$ is the annihilation operator for the eigenstates.

\subsection{Electron-phonon coupling Hamiltonian}

The coupling to the in-plane modes (here LO,TO) is obtained
by assuming that when the equilibrium nearest-neighbor distance 
$a$ is perturbed by $\delta a$, then
$\gamma_0\rightarrow \gamma_0+\gamma_0'\delta a$.
For BLG this leads to an electron-phonon 
coupling Hamiltonian of the form\cite{andoreview05,castroneto09,ando09}
\begin{equation}\begin{split} \label{e.lteph}
\hat{H}_{e-ph}^{(LT)}
=&\frac{\gamma_0'}{a}\sum_{\vec{R}}\sum_{j=1,2,3}
  \psi_{A1}^\dagger(\vec{R})\psi_{B1}(\vec{R}+\vec{\delta}_j) \\
  &\times
  \vec{\delta}_j\cdot[\vec{u}_{A1}(\vec{R})
  -\vec{u}_{B1}(\vec{R}+\vec{\delta}_j)] \\
&+\frac{\gamma_0'}{a}\sum_{\vec{R}_s}\sum_{j=1,2,3}
  \psi_{A2}^\dagger(\vec{R}_s)\psi_{B2}(\vec{R}_s+\vec{\delta}_j) \\
  &\times
  \vec{\delta}_j\cdot[\vec{u}_{A2}(\vec{R}_s)
  -\vec{u}_{B2}(\vec{R}_s+\vec{\delta}_j)] 
+h.c.,
\end{split}\end{equation}
where $\vec{R}$ sums over the 
A1 lattice sites, $\vec{R}_s$ are the A2 sites,
and where $\vec{u}_{\delta\sigma}$ is the in-plane displacement of
atom $\delta=A,B$ in layer $\sigma=1,2$.
Here we may insert the Fourier transformations
\begin{subequations}
\begin{align}
\psi_{A\sigma}(\vec{R}) & = \frac{1}{\sqrt{N}} 
\sum_{\vec{k}} e^{i\vec{k}\cdot\vec{R}} a_{\sigma\vec{k}}, \quad
\sigma=1,2, \label{e.ft1} \\
\vec{u}_{A\sigma}(\vec{R}) & = \frac{1}{\sqrt{N}} 
\sum_{\vec{q}} e^{i\vec{q}\cdot\vec{R}} \vec{U}_{A\sigma\vec{q}}, \quad
\sigma=1,2, \label{e.ft2}
\end{align}
\end{subequations}
where $N$ is the number of unit cells,
and similar ones with $A\rightarrow B$ and 
$a_{\sigma\vec{k}}\rightarrow b_{\sigma\vec{k}}$.
Then, calculating the wave vectors relative to 
$\vec{K}$ or $\vec{K}'$ and assuming $q\ll|\vec{K}|=4\pi/(3\sqrt{3}a)$,
we obtain\cite{andoreview05} 
\begin{equation}\begin{split}
&\hat{H}_{e-ph}^{(LT)} =
\frac{3\gamma_0'}{2}\frac{1}{\sqrt{N}} 
\sum_{\vec{p}}\sum_{\vec{q}}
\Psi_{\vec{p}+\vec{q}}^\dagger \times \\
&
\left(
\begin{matrix}
0     & U^y_{1\vec{q}}\mp i U^x_{1\vec{q}}   & 0     &  0    \\
U^y_{1\vec{q}}\pm i U^x_{1\vec{q}}     & 0   & 0     &  0    \\
0     & 0     & 0      &  U^y_{2\vec{q}}\pm i U^x_{2\vec{q}} \\
0     & 0     &  U^y_{2\vec{q}}\mp i U^x_{2\vec{q}}     &  0    
\end{matrix}
\right)
\Psi_{\vec{p}},
\end{split}\end{equation}
where 
$\vec{U}_{\sigma\vec{q}} 
= \vec{U}_{A\sigma\vec{q}} - \vec{U}_{B\sigma\vec{q}}$
($\sigma=1,2$)
is the relative displacement vector between A and B atoms.

For the ZO optical modes, which can modulate the
equilibrium distance $d$ between the A1 and B2 atoms
such that
$\gamma_1\rightarrow \gamma_1+\gamma_1'\delta d$, 
the coupling Hamiltonian is of the form
\begin{equation}\begin{split} \label{e.zaeph}
\hat{H}_{e-ph}^{(ZO)}
=&\frac{\gamma_1'}{d}\sum_{\vec{R}}\sum_{j=1,2,3}
[
  \psi_{A1}^\dagger(\vec{R})\psi_{B2}(\vec{R}_d) \\
  &+\psi_{B2}^\dagger(\vec{R}_d)\psi_{A1}(\vec{R})
]
[h_{B2}(\vec{R}_d)-h_{A1}(\vec{R})].
\end{split}\end{equation}
Here $h_{\delta\sigma}$ is the out-of-plane displacement of 
atom $\delta\sigma$ and $\vec{R}_d$ is the site of the B2
atom above $\vec{R}$.
Inserting Eq.\ (\ref{e.ft1}) and
\begin{equation}\begin{split} \label{e.ft3}
h_{\delta\sigma}(\vec{R}) = \frac{1}{\sqrt{N}}\sum_{\vec{q}} 
e^{i\vec{q}\cdot\vec{R}}H_{\delta\sigma\vec{q}}
\end{split}\end{equation}
we find
\begin{equation}\begin{split}
\hat{H}_{e-ph}^{(ZO)} &=
\gamma_1'\frac{1}{\sqrt{N}}
\sum_{\vec{p}}\sum_{\vec{q}}
\Psi_{\vec{p}+\vec{q}}^\dagger
\left(
\begin{matrix}
0     & 0    & H_{\vec{q}}    &  0    \\
0     & 0    & 0    &  0    \\
H_{\vec{q}}     & 0    & 0    &  0 \\
0     & 0    & 0    &  0    
\end{matrix}
\right)
\Psi_{\vec{p}},
\end{split}\end{equation}
where $H_{\vec{q}}=H_{B2\vec{q}}-H_{A1\vec{q}}$.

These results must still be expressed in the
low-energy eigenbasis by using the expansion
$\Psi_{\vec{k}}=\sum_{\alpha}\Phi_\alpha(\vec{k})c_{\vec{k}\alpha}$,
and where $\alpha$ sums only over the low-energy bands (${\rm c,v}$),
keeping only terms to first order in $\hbar v k/\gamma_1$.
Also, a transformation of the form
\begin{equation}\begin{split}
(\vec{U}_{\delta\sigma\vec{q}})_\zeta & = \sum_\gamma X_{\delta\sigma\zeta,\gamma}(\vec{q}) 
\tilde{\vec{U}}_{\vec{q}\gamma} \\
H_{\delta\sigma\vec{q}} & = \sum_\gamma X_{\delta\sigma z,\gamma}(\vec{q}) 
\tilde{\vec{U}}_{\vec{q}\gamma} \\
\end{split}\end{equation}
where $\zeta=x,y$,
has to be introduced, with appropriate models 
for the matrix $X(\vec{q})$ as discussed below.
Finally, the phonons are quantized with
$\tilde{\vec{U}}_{\vec{q}\gamma} = \sqrt{\hbar^2/(2\Omega_\gamma(\vec{q}))}
(b_{\vec{q}\gamma} + b_{-\vec{q}\gamma}^\dagger )$,
where $\Omega_\gamma(\vec{q})$ is the energy of the phonon mode.
This leads to the form
\begin{equation}\begin{split} \label{e.ephham}
\hat{H}_{e-ph}
&=
\sum_{\vec{q}\gamma}
\sum_{\vec{k}\alpha}
\sum_{\vec{p}\beta}
M_{\vec{p}\vec{k},\vec{q}}^{\beta\alpha,\gamma}
c_{\vec{p}\beta}^\dagger c_{\vec{k}\alpha}
(b_{\vec{q}\gamma} + b_{-\vec{q}\gamma}^\dagger)
\end{split}\end{equation}
such that the matrix elements
$M_{\vec{p}\vec{k},\vec{q}}^{\beta\alpha,\gamma}$ and thus the
coupling constants may be identified.
For MLG the derivation is similar.


\subsection{Coupling constants for simple intrinsic phonon models}

Next we consider some simple models for long-wavelength optical
phonons.\cite{suzuura02,andoreview05,tsedassarma07}  
We do this by
specifying the form of the transformation matrix $X(\vec{q})$ for the
different phonon branches, which 
should otherwise be obtained by diagonalizing the phonon 
Hamiltonian.  The matrix must be normalized as
$\sum_{\delta\sigma\zeta} X_{\delta\sigma\zeta,\mu}^*(\vec{q})M_{\delta\sigma}
X_{\delta\sigma\zeta,\nu} = \delta_{\mu\nu}$,
where $M_{\delta\sigma}$ is the mass of atom $\delta\sigma$ in
the unit cell. In graphene $M_{\delta \sigma}=M_C$, the mass of a
carbon atom.  Below we use the vector notation
$(\vec{X}_{\delta\sigma,\gamma})_{\zeta}=X_{\delta\sigma\zeta,\mu}$.

Let us first consider a model for the long-wavelength in-plane (LT)
optical phonons in MLG, for which the LO and TO branches are nearly
degenerate. To describe the characteristic opposite-phase motion of the 
A and B sublattices we choose
\begin{equation}\begin{split} \label{e.monoloto}
\vec{X}_{A1,LT}(\vec{q}) &= \frac{1}{\sqrt{2M_C}}\hat{\vec{a}}(\vec{q}),
\quad
\vec{X}_{B1,LT}(\vec{q}) = \frac{-1}{\sqrt{2M_C}}\hat{\vec{a}}(\vec{q}),
\end{split}\end{equation}
where
$\hat{\vec{a}}(\vec{q})=\hat{\vec{q}}$ for the LO and
$\hat{\vec{a}}(\vec{q})=\hat{\vec{z}}\times\hat{\vec{q}}$
for the TO branch, $\hat{\vec{z}}$ being normal to the plane. 
This yields\cite{tsedassarma09}
\begin{equation} \label{e.mlltcc}
w_{\vec{k}\vec{p},\vec{q}}^{\alpha\beta,LT}
=
\frac{9(\gamma_0')^2\hbar^2}{2M\Omega_{LT}}
\frac{1}{2}(1-s_\alpha s_\beta 
\cos(\phi_{\vec{k}} + \phi_{\vec{p}} -2\phi_{\hat{\vec{a}}})),
\end{equation}
where $M=2M_CN$.

In BLG there are four nearly-degenerate 
LT branches. For them we choose similarly
\begin{equation}\begin{split} \label{e.biloto}
\vec{X}_{A1,LT}(\vec{q}) &= \frac{1}{2\sqrt{M_C}}\hat{\vec{a}}(\vec{q}),
\quad
\vec{X}_{B1,LT}(\vec{q}) = \frac{-1}{2\sqrt{M_C}}\hat{\vec{a}}(\vec{q}),
\\
\vec{X}_{A2,LT}(\vec{q}) &= \frac{\pm 1}{2\sqrt{M_C}}\hat{\vec{a}}(\vec{q}),
\quad
\vec{X}_{B2,LT}(\vec{q}) = \frac{\mp 1}{2\sqrt{M_C}}\hat{\vec{a}}(\vec{q}).
\end{split}\end{equation}
According to these, for both LO and TO type modes the 
atoms in layers 1 and 2 can move
either in phase (upper signs) or in opposite phases (lower signs).  
We find
\begin{equation}\begin{split} \label{e.blltcc}
w_{\vec{k}\vec{p},\vec{q}}^{\alpha\beta,LT}
&=
\frac{9(\gamma_0')^2\hbar^2}{2M\Omega_{LT}}
\frac{1}{2}\frac{(\hbar v)^2}{\gamma_1^2}
\bigg\{
k^2 + p^2 \\
+ & 2kp
\left[
\pm\cos(\phi_{\vec{k}\vec{p}})
- s_{\alpha}s_{\beta}\cos(\phi_{\vec{k}\hat{\vec{a}}} + \phi_{\vec{p}\hat{\vec{a}}})
\right] \\
\mp & s_{\alpha}s_{\beta}
\left[
k^2\cos(2\phi_{\vec{p}\hat{\vec{a}}})
+
p^2\cos(2\phi_{\vec{k}\hat{\vec{a}}})
\right]
\bigg\}
\end{split}\end{equation} 
with $M=4M_CN$ and $\phi_{\vec{k}\vec{p}}=\phi_{\vec{k}}-\phi_{\vec{p}}$. 
To model the ZO modes of BLG, we choose
\begin{equation}\begin{split} \label{e.bizo}
\vec{X}_{A1,ZO}(\vec{q}) &= \frac{1}{2\sqrt{M_C}}\hat{\vec{z}},
\quad
\vec{X}_{B1,ZO}(\vec{q}) = \frac{- 1}{2\sqrt{M_C}}\hat{\vec{z}},
\\
\vec{X}_{A2,ZO}(\vec{q}) &= \frac{ 1}{2\sqrt{M_C}}\hat{\vec{z}},
\quad
\vec{X}_{B2,ZO}(\vec{q}) = \frac{-1}{2\sqrt{M_C}}\hat{\vec{z}}.
\end{split}\end{equation}
where the signs amplitudes of the A2 and B1 displacements are 
unimportant in our approximation, which neglects any coupling 
between them. This yields the coupling constants
\begin{equation}
w_{\vec{k}\vec{p},\vec{q}}^{\alpha\beta,ZO}
=
\frac{2(\gamma_1')^2\hbar^2}{M\Omega_{ZO}}
\delta_{\alpha\beta}
(\hbar v/\gamma_1)^2kp,
\end{equation}
with $M=4M_CN$. 
This form is particularly simple, because there is no angular
dependence, and no interband coupling. In fact, for $k=p$
($q\rightarrow 0$) the result may be obtained from the continuum
description by modulating $\gamma_1$,\emph{i.e.} the
effective electron mass $\gamma_1/(2v^2)$.



\begin{thebibliography}{52}
\expandafter\ifx\csname natexlab\endcsname\relax\def\natexlab#1{#1}\fi
\expandafter\ifx\csname bibnamefont\endcsname\relax
  \def\bibnamefont#1{#1}\fi
\expandafter\ifx\csname bibfnamefont\endcsname\relax
  \def\bibfnamefont#1{#1}\fi
\expandafter\ifx\csname citenamefont\endcsname\relax
  \def\citenamefont#1{#1}\fi
\expandafter\ifx\csname url\endcsname\relax
  \def\url#1{\texttt{#1}}\fi
\expandafter\ifx\csname urlprefix\endcsname\relax\def\urlprefix{URL }\fi
\providecommand{\bibinfo}[2]{#2}
\providecommand{\eprint}[2][]{\url{#2}}

\bibitem[{\citenamefont{Allen}(1987)}]{allen87}
\bibinfo{author}{\bibfnamefont{P.~B.} \bibnamefont{Allen}},
  \bibinfo{journal}{Phys. Rev. Lett.} \textbf{\bibinfo{volume}{59}},
  \bibinfo{pages}{1460} (\bibinfo{year}{1987}).

\bibitem[{\citenamefont{Wellstood et~al.}(1994)\citenamefont{Wellstood, Urbina,
  and Clarke}}]{wellstood94}
\bibinfo{author}{\bibfnamefont{F.~C.} \bibnamefont{Wellstood}},
  \bibinfo{author}{\bibfnamefont{C.}~\bibnamefont{Urbina}}, \bibnamefont{and}
  \bibinfo{author}{\bibfnamefont{J.}~\bibnamefont{Clarke}},
  \bibinfo{journal}{Phys. Rev. B} \textbf{\bibinfo{volume}{49}},
  \bibinfo{pages}{5942} (\bibinfo{year}{1994}).

\bibitem[{\citenamefont{Giazotto et~al.}(2006)\citenamefont{Giazotto,
  Heikkil\"{a}, Luukanen, Savin, and Pekola}}]{giazotto06}
\bibinfo{author}{\bibfnamefont{F.}~\bibnamefont{Giazotto}},
  \bibinfo{author}{\bibfnamefont{T.~T.} \bibnamefont{Heikkil\"{a}}},
  \bibinfo{author}{\bibfnamefont{A.}~\bibnamefont{Luukanen}},
  \bibinfo{author}{\bibfnamefont{A.~M.} \bibnamefont{Savin}}, \bibnamefont{and}
  \bibinfo{author}{\bibfnamefont{J.~P.} \bibnamefont{Pekola}},
  \bibinfo{journal}{Rev. Mod. Phys.} \textbf{\bibinfo{volume}{78}},
  \bibinfo{eid}{217} (\bibinfo{year}{2006}).

\bibitem[{\citenamefont{Hekking et~al.}(2008)\citenamefont{Hekking, Niskanen,
  and Pekola}}]{hekking08}
\bibinfo{author}{\bibfnamefont{F.~W.~J.} \bibnamefont{Hekking}},
  \bibinfo{author}{\bibfnamefont{A.~O.} \bibnamefont{Niskanen}},
  \bibnamefont{and} \bibinfo{author}{\bibfnamefont{J.~P.}
  \bibnamefont{Pekola}}, \bibinfo{journal}{Phys. Rev. B}
  \textbf{\bibinfo{volume}{77}}, \bibinfo{eid}{033401} (\bibinfo{year}{2008}).

\bibitem[{\citenamefont{Mahan}(2000)}]{Mahan3rd}
\bibinfo{author}{\bibfnamefont{G.~D.} \bibnamefont{Mahan}},
  \emph{\bibinfo{title}{Many-Particle Physics, 3rd ed.}}
  (\bibinfo{publisher}{Kluwer Academic/Plenum Publishers},
  \bibinfo{address}{New York}, \bibinfo{year}{2000}).

\bibitem[{\citenamefont{Sergeev and Mitin}(2000)}]{sergeev00}
\bibinfo{author}{\bibfnamefont{A.}~\bibnamefont{Sergeev}} \bibnamefont{and}
  \bibinfo{author}{\bibfnamefont{V.}~\bibnamefont{Mitin}},
  \bibinfo{journal}{Phys. Rev. B} \textbf{\bibinfo{volume}{61}},
  \bibinfo{pages}{6041} (\bibinfo{year}{2000}).

\bibitem[{\citenamefont{Karvonen et~al.}(2005)\citenamefont{Karvonen, Taskinen,
  and Maasilta}}]{karvonen05}
\bibinfo{author}{\bibfnamefont{J.~T.} \bibnamefont{Karvonen}},
  \bibinfo{author}{\bibfnamefont{L.~J.} \bibnamefont{Taskinen}},
  \bibnamefont{and} \bibinfo{author}{\bibfnamefont{I.~J.}
  \bibnamefont{Maasilta}}, \bibinfo{journal}{Phys. Rev. B}
  \textbf{\bibinfo{volume}{72}}, \bibinfo{pages}{012302}
  (\bibinfo{year}{2005}).

\bibitem[{\citenamefont{Kubakaddi}(2009)}]{kubakaddi09}
\bibinfo{author}{\bibfnamefont{S.~S.} \bibnamefont{Kubakaddi}},
  \bibinfo{journal}{Phys. Rev. B} \textbf{\bibinfo{volume}{79}},
  \bibinfo{pages}{075417} (\bibinfo{year}{2009}).

\bibitem[{\citenamefont{Bistritzer and
  MacDonald}(2009{\natexlab{a}})}]{bistritzer09}
\bibinfo{author}{\bibfnamefont{R.}~\bibnamefont{Bistritzer}} \bibnamefont{and}
  \bibinfo{author}{\bibfnamefont{A.~H.} \bibnamefont{MacDonald}},
  \bibinfo{journal}{Phys. Rev. Lett.} \textbf{\bibinfo{volume}{102}},
  \bibinfo{eid}{206410} (\bibinfo{year}{2009}{\natexlab{a}}).

\bibitem[{\citenamefont{Tse and {Das Sarma}}(2009)}]{tsedassarma09}
\bibinfo{author}{\bibfnamefont{W.-K.} \bibnamefont{Tse}} \bibnamefont{and}
  \bibinfo{author}{\bibfnamefont{S.}~\bibnamefont{{Das Sarma}}},
  \bibinfo{journal}{Phys. Rev. B} \textbf{\bibinfo{volume}{79}},
  \bibinfo{eid}{235406} (\bibinfo{year}{2009}).

\bibitem[{\citenamefont{Novoselov et~al.}(2004)\citenamefont{Novoselov, Geim,
  Morozov, Jiang, Zhang, Dubonos, Grigorieva, and Firsov}}]{novoselov04}
\bibinfo{author}{\bibfnamefont{K.~S.} \bibnamefont{Novoselov}},
  \bibinfo{author}{\bibfnamefont{A.~K.} \bibnamefont{Geim}},
  \bibinfo{author}{\bibfnamefont{S.~V.} \bibnamefont{Morozov}},
  \bibinfo{author}{\bibfnamefont{D.}~\bibnamefont{Jiang}},
  \bibinfo{author}{\bibfnamefont{Y.}~\bibnamefont{Zhang}},
  \bibinfo{author}{\bibfnamefont{S.~V.} \bibnamefont{Dubonos}},
  \bibinfo{author}{\bibfnamefont{I.~V.} \bibnamefont{Grigorieva}},
  \bibnamefont{and} \bibinfo{author}{\bibfnamefont{A.~A.}
  \bibnamefont{Firsov}}, \bibinfo{journal}{Science}
  \textbf{\bibinfo{volume}{306}}, \bibinfo{pages}{666} (\bibinfo{year}{2004}).

\bibitem[{\citenamefont{{Castro Neto} et~al.}(2009)\citenamefont{{Castro Neto},
  Guinea, Peres, Novoselov, and Geim}}]{castroneto09}
\bibinfo{author}{\bibfnamefont{A.~H.} \bibnamefont{{Castro Neto}}},
  \bibinfo{author}{\bibfnamefont{F.}~\bibnamefont{Guinea}},
  \bibinfo{author}{\bibfnamefont{N.~M.~R.} \bibnamefont{Peres}},
  \bibinfo{author}{\bibfnamefont{K.~S.} \bibnamefont{Novoselov}},
  \bibnamefont{and} \bibinfo{author}{\bibfnamefont{A.~K.} \bibnamefont{Geim}},
  \bibinfo{journal}{Rev. Mod. Phys.} \textbf{\bibinfo{volume}{81}},
  \bibinfo{eid}{109} (\bibinfo{year}{2009}).

\bibitem[{\citenamefont{McCann and Fal'ko}(2006)}]{mccann06}
\bibinfo{author}{\bibfnamefont{E.}~\bibnamefont{McCann}} \bibnamefont{and}
  \bibinfo{author}{\bibfnamefont{V.~I.} \bibnamefont{Fal'ko}},
  \bibinfo{journal}{Phys. Rev. Lett.} \textbf{\bibinfo{volume}{96}},
  \bibinfo{eid}{086805} (\bibinfo{year}{2006}).

\bibitem[{\citenamefont{Hwang and {Das Sarma}}(2008)}]{hwang08}
\bibinfo{author}{\bibfnamefont{E.~H.} \bibnamefont{Hwang}} \bibnamefont{and}
  \bibinfo{author}{\bibfnamefont{S.}~\bibnamefont{{Das Sarma}}},
  \bibinfo{journal}{Phys. Rev. B} \textbf{\bibinfo{volume}{77}},
  \bibinfo{pages}{115449} (\bibinfo{year}{2008}).

\bibitem[{\citenamefont{Fratini and Guinea}(2008)}]{fratini08}
\bibinfo{author}{\bibfnamefont{S.}~\bibnamefont{Fratini}} \bibnamefont{and}
  \bibinfo{author}{\bibfnamefont{F.}~\bibnamefont{Guinea}},
  \bibinfo{journal}{Phys. Rev. B} \textbf{\bibinfo{volume}{77}},
  \bibinfo{pages}{195415} (\bibinfo{year}{2008}).

\bibitem[{\citenamefont{Meric et~al.}(2008)\citenamefont{Meric, Han, Young,
  Ozyilmaz, Kim, and Shepard}}]{meric08}
\bibinfo{author}{\bibfnamefont{I.}~\bibnamefont{Meric}},
  \bibinfo{author}{\bibfnamefont{M.~Y.} \bibnamefont{Han}},
  \bibinfo{author}{\bibfnamefont{A.~F.} \bibnamefont{Young}},
  \bibinfo{author}{\bibfnamefont{B.}~\bibnamefont{Ozyilmaz}},
  \bibinfo{author}{\bibfnamefont{P.}~\bibnamefont{Kim}}, \bibnamefont{and}
  \bibinfo{author}{\bibfnamefont{K.~L.} \bibnamefont{Shepard}},
  \bibinfo{journal}{Nature Nanotech.} \textbf{\bibinfo{volume}{3}},
  \bibinfo{pages}{854} (\bibinfo{year}{2008}).

\bibitem[{\citenamefont{Mariani and von Oppen}(2008)}]{mariani08}
\bibinfo{author}{\bibfnamefont{E.}~\bibnamefont{Mariani}} \bibnamefont{and}
  \bibinfo{author}{\bibfnamefont{F.}~\bibnamefont{von Oppen}},
  \bibinfo{journal}{Phys. Rev. Lett.} \textbf{\bibinfo{volume}{100}},
  \bibinfo{eid}{076801} (\bibinfo{year}{2008}), \bibinfo{note}{{Phys. Rev.
  Lett. \textbf{100}, 249901(E) (2008)}}.

\bibitem[{\citenamefont{Chen et~al.}(2009)\citenamefont{Chen, Jang, Bao, Lau,
  and Dames}}]{chen09}
\bibinfo{author}{\bibfnamefont{Z.}~\bibnamefont{Chen}},
  \bibinfo{author}{\bibfnamefont{W.}~\bibnamefont{Jang}},
  \bibinfo{author}{\bibfnamefont{W.}~\bibnamefont{Bao}},
  \bibinfo{author}{\bibfnamefont{C.~N.} \bibnamefont{Lau}}, \bibnamefont{and}
  \bibinfo{author}{\bibfnamefont{C.}~\bibnamefont{Dames}},
  \bibinfo{journal}{Appl. Phys. Lett.} \textbf{\bibinfo{volume}{95}},
  \bibinfo{eid}{161910} (\bibinfo{year}{2009}).

\bibitem[{\citenamefont{Bistritzer and
  MacDonald}(2009{\natexlab{b}})}]{bistritzer09b}
\bibinfo{author}{\bibfnamefont{R.}~\bibnamefont{Bistritzer}} \bibnamefont{and}
  \bibinfo{author}{\bibfnamefont{A.~H.} \bibnamefont{MacDonald}},
  \bibinfo{journal}{Phys. Rev. B} \textbf{\bibinfo{volume}{80}},
  \bibinfo{pages}{085109} (\bibinfo{year}{2009}{\natexlab{b}}).

\bibitem[{\citenamefont{Kabanov and Alexandrov}(2008)}]{kabanov08}
\bibinfo{author}{\bibfnamefont{V.~V.} \bibnamefont{Kabanov}} \bibnamefont{and}
  \bibinfo{author}{\bibfnamefont{A.~S.} \bibnamefont{Alexandrov}},
  \bibinfo{journal}{Phys. Rev. B} \textbf{\bibinfo{volume}{78}},
  \bibinfo{pages}{174514} (\bibinfo{year}{2008}).

\bibitem[{\citenamefont{Slonczewski and Weiss}(1958)}]{slonczewski58}
\bibinfo{author}{\bibfnamefont{J.~C.} \bibnamefont{Slonczewski}}
  \bibnamefont{and} \bibinfo{author}{\bibfnamefont{P.~R.} \bibnamefont{Weiss}},
  \bibinfo{journal}{Phys. Rev.} \textbf{\bibinfo{volume}{109}},
  \bibinfo{pages}{272} (\bibinfo{year}{1958}).

\bibitem[{\citenamefont{McClure}(1957)}]{mcclure57}
\bibinfo{author}{\bibfnamefont{J.~W.} \bibnamefont{McClure}},
  \bibinfo{journal}{Phys. Rev.} \textbf{\bibinfo{volume}{108}},
  \bibinfo{pages}{612} (\bibinfo{year}{1957}).

\bibitem[{\citenamefont{Dresselhaus and Dresselhaus}(2002)}]{dresselhaus02}
\bibinfo{author}{\bibfnamefont{M.~S.} \bibnamefont{Dresselhaus}}
  \bibnamefont{and}
  \bibinfo{author}{\bibfnamefont{G.}~\bibnamefont{Dresselhaus}},
  \bibinfo{journal}{Adv. Phys.} \textbf{\bibinfo{volume}{51}},
  \bibinfo{pages}{1} (\bibinfo{year}{2002}).

\bibitem[{\citenamefont{Nilsson et~al.}(2008)\citenamefont{Nilsson, {Castro
  Neto}, Guinea, and Peres}}]{nilsson08}
\bibinfo{author}{\bibfnamefont{J.}~\bibnamefont{Nilsson}},
  \bibinfo{author}{\bibfnamefont{A.~H.} \bibnamefont{{Castro Neto}}},
  \bibinfo{author}{\bibfnamefont{F.}~\bibnamefont{Guinea}}, \bibnamefont{and}
  \bibinfo{author}{\bibfnamefont{N.~M.~R.} \bibnamefont{Peres}},
  \bibinfo{journal}{Phys. Rev. B} \textbf{\bibinfo{volume}{78}},
  \bibinfo{eid}{045405} (\bibinfo{year}{2008}).

\bibitem[{\citenamefont{Suzuura and Ando}(2008)}]{suzuura08}
\bibinfo{author}{\bibfnamefont{H.}~\bibnamefont{Suzuura}} \bibnamefont{and}
  \bibinfo{author}{\bibfnamefont{T.}~\bibnamefont{Ando}}, \bibinfo{journal}{J.
  Phys. Soc. Jpn.} \textbf{\bibinfo{volume}{77}}, \bibinfo{pages}{044703}
  (\bibinfo{year}{2008}).

\bibitem[{\citenamefont{Rana et~al.}(2009)\citenamefont{Rana, George, Strait,
  Dawlaty, Shivaraman, Chandrashekhar, and Spencer}}]{rana09}
\bibinfo{author}{\bibfnamefont{F.}~\bibnamefont{Rana}},
  \bibinfo{author}{\bibfnamefont{P.~A.} \bibnamefont{George}},
  \bibinfo{author}{\bibfnamefont{J.~H.} \bibnamefont{Strait}},
  \bibinfo{author}{\bibfnamefont{J.}~\bibnamefont{Dawlaty}},
  \bibinfo{author}{\bibfnamefont{S.}~\bibnamefont{Shivaraman}},
  \bibinfo{author}{\bibfnamefont{M.}~\bibnamefont{Chandrashekhar}},
  \bibnamefont{and} \bibinfo{author}{\bibfnamefont{M.~G.}
  \bibnamefont{Spencer}}, \bibinfo{journal}{Phys. Rev. B}
  \textbf{\bibinfo{volume}{79}}, \bibinfo{pages}{115447}
  (\bibinfo{year}{2009}).

\bibitem[{\citenamefont{Suzuura and Ando}(2002)}]{suzuura02}
\bibinfo{author}{\bibfnamefont{H.}~\bibnamefont{Suzuura}} \bibnamefont{and}
  \bibinfo{author}{\bibfnamefont{T.}~\bibnamefont{Ando}},
  \bibinfo{journal}{Phys. Rev. B} \textbf{\bibinfo{volume}{65}},
  \bibinfo{pages}{235412} (\bibinfo{year}{2002}).

\bibitem[{\citenamefont{Rotkin et~al.}(2009)\citenamefont{Rotkin, Perebeinos,
  Petrov, and Avouris}}]{rotkin09}
\bibinfo{author}{\bibfnamefont{S.~V.} \bibnamefont{Rotkin}},
  \bibinfo{author}{\bibfnamefont{V.}~\bibnamefont{Perebeinos}},
  \bibinfo{author}{\bibfnamefont{A.~G.} \bibnamefont{Petrov}},
  \bibnamefont{and} \bibinfo{author}{\bibfnamefont{P.}~\bibnamefont{Avouris}},
  \bibinfo{journal}{Nano Lett.} \textbf{\bibinfo{volume}{9}},
  \bibinfo{pages}{1850} (\bibinfo{year}{2009}).

\bibitem[{\citenamefont{Mohr et~al.}(2007)\citenamefont{Mohr, Maultzsch,
  Dobard\v{z}i\'{c}, Reich, Milo\v{s}evi\'{c}, Damnjanovi\'{c}, Bosak, Krisch,
  and Thomsen}}]{mohr07}
\bibinfo{author}{\bibfnamefont{M.}~\bibnamefont{Mohr}},
  \bibinfo{author}{\bibfnamefont{J.}~\bibnamefont{Maultzsch}},
  \bibinfo{author}{\bibfnamefont{E.}~\bibnamefont{Dobard\v{z}i\'{c}}},
  \bibinfo{author}{\bibfnamefont{S.}~\bibnamefont{Reich}},
  \bibinfo{author}{\bibfnamefont{I.}~\bibnamefont{Milo\v{s}evi\'{c}}},
  \bibinfo{author}{\bibfnamefont{M.}~\bibnamefont{Damnjanovi\'{c}}},
  \bibinfo{author}{\bibfnamefont{A.}~\bibnamefont{Bosak}},
  \bibinfo{author}{\bibfnamefont{M.}~\bibnamefont{Krisch}}, \bibnamefont{and}
  \bibinfo{author}{\bibfnamefont{C.}~\bibnamefont{Thomsen}},
  \bibinfo{journal}{Phys. Rev. B} \textbf{\bibinfo{volume}{76}},
  \bibinfo{eid}{035439} (\bibinfo{year}{2007}).

\bibitem[{\citenamefont{Yan et~al.}(2008)\citenamefont{Yan, Ruan, and
  Chou}}]{yan08}
\bibinfo{author}{\bibfnamefont{J.-A.} \bibnamefont{Yan}},
  \bibinfo{author}{\bibfnamefont{W.~Y.} \bibnamefont{Ruan}}, \bibnamefont{and}
  \bibinfo{author}{\bibfnamefont{M.~Y.} \bibnamefont{Chou}},
  \bibinfo{journal}{Phys. Rev. B} \textbf{\bibinfo{volume}{77}},
  \bibinfo{eid}{125401} (\bibinfo{year}{2008}).

\bibitem[{\citenamefont{Michel and Verberck}(2008)}]{michel08}
\bibinfo{author}{\bibfnamefont{K.~H.} \bibnamefont{Michel}} \bibnamefont{and}
  \bibinfo{author}{\bibfnamefont{B.}~\bibnamefont{Verberck}},
  \bibinfo{journal}{Phys. Rev. B} \textbf{\bibinfo{volume}{78}},
  \bibinfo{eid}{085424} (\bibinfo{year}{2008}).

\bibitem[{\citenamefont{Malard et~al.}(2009)\citenamefont{Malard,
  Guimar{\~a}es, Mafra, Mazzoni, and Jorio}}]{malard09}
\bibinfo{author}{\bibfnamefont{L.~M.} \bibnamefont{Malard}},
  \bibinfo{author}{\bibfnamefont{M.~H.~D.} \bibnamefont{Guimar{\~a}es}},
  \bibinfo{author}{\bibfnamefont{D.~L.} \bibnamefont{Mafra}},
  \bibinfo{author}{\bibfnamefont{M.~S.~C.} \bibnamefont{Mazzoni}},
  \bibnamefont{and} \bibinfo{author}{\bibfnamefont{A.}~\bibnamefont{Jorio}},
  \bibinfo{journal}{Phys. Rev. B} \textbf{\bibinfo{volume}{79}},
  \bibinfo{eid}{125426} (\bibinfo{year}{2009}).

\bibitem[{\citenamefont{{Viola Kusminskiy} et~al.}(2009)\citenamefont{{Viola
  Kusminskiy}, Campbell, and {Castro Neto}}}]{kuminsky09}
\bibinfo{author}{\bibfnamefont{S.}~\bibnamefont{{Viola Kusminskiy}}},
  \bibinfo{author}{\bibfnamefont{D.~K.} \bibnamefont{Campbell}},
  \bibnamefont{and} \bibinfo{author}{\bibfnamefont{A.~H.} \bibnamefont{{Castro
  Neto}}}, \bibinfo{journal}{Phys. Rev. B} \textbf{\bibinfo{volume}{80}},
  \bibinfo{pages}{035401} (\bibinfo{year}{2009}).

\bibitem[{\citenamefont{Ando}(2005)}]{andoreview05}
\bibinfo{author}{\bibfnamefont{T.}~\bibnamefont{Ando}}, \bibinfo{journal}{J.
  Phys. Soc. Jpn.} \textbf{\bibinfo{volume}{74}}, \bibinfo{pages}{777}
  (\bibinfo{year}{2005}).

\bibitem[{\citenamefont{Tse and Das~Sarma}(2007)}]{tsedassarma07}
\bibinfo{author}{\bibfnamefont{W.-K.} \bibnamefont{Tse}} \bibnamefont{and}
  \bibinfo{author}{\bibfnamefont{S.}~\bibnamefont{Das~Sarma}},
  \bibinfo{journal}{Phys. Rev. Lett.} \textbf{\bibinfo{volume}{99}},
  \bibinfo{pages}{236802} (\bibinfo{year}{2007}).

\bibitem[{\citenamefont{Ando and Koshino}(2009)}]{ando09}
\bibinfo{author}{\bibfnamefont{T.}~\bibnamefont{Ando}} \bibnamefont{and}
  \bibinfo{author}{\bibfnamefont{M.}~\bibnamefont{Koshino}},
  \bibinfo{journal}{J. Phys. Soc. Jpn.} \textbf{\bibinfo{volume}{78}},
  \bibinfo{pages}{034709} (\bibinfo{year}{2009}).

\bibitem[{\citenamefont{Wang and Mahan}(1972)}]{wang72}
\bibinfo{author}{\bibfnamefont{S.~Q.} \bibnamefont{Wang}} \bibnamefont{and}
  \bibinfo{author}{\bibfnamefont{G.~D.} \bibnamefont{Mahan}},
  \bibinfo{journal}{Phys. Rev. B} \textbf{\bibinfo{volume}{6}},
  \bibinfo{pages}{4517} (\bibinfo{year}{1972}).

\bibitem[{\citenamefont{Fischetti et~al.}(2001)\citenamefont{Fischetti,
  Neumayer, and Cartier}}]{fischetti01}
\bibinfo{author}{\bibfnamefont{M.~V.} \bibnamefont{Fischetti}},
  \bibinfo{author}{\bibfnamefont{D.~A.} \bibnamefont{Neumayer}},
  \bibnamefont{and} \bibinfo{author}{\bibfnamefont{E.~A.}
  \bibnamefont{Cartier}}, \bibinfo{journal}{J. Appl. Phys.}
  \textbf{\bibinfo{volume}{90}}, \bibinfo{pages}{4587} (\bibinfo{year}{2001}).

\bibitem[{pap()}]{papanote}
\bibinfo{note}{Unlike acoustic phonons or the remote optical phonons of a
  substrate, the intrinsic optical modes modulate the onsite energies only
  weakly. We have checked this with the tight-binding parametrization of
  D.~A.~Papaconstantopoulos and M.~J.~Mehl, J.\ Phys.\ Condens.\ Matter,
  \textbf{15}, R413 (2003), which yields onsite shifts roughly an order of
  magnitude smaller than those of the hopping integrals for a perturbation
  corresponding to a $\Gamma$ point LO/TO mode.}

\bibitem[{zon()}]{zonote}
\bibinfo{note}{By modeling the ZO modes of bilayer as a mode where the A1 and
  B2 atoms move in opposite phases, thus modulating $\gamma_1$ (\emph{i.e.} the
  effective mass), we find a coupling constant of the form
  $w_{\vec{k}\vec{p},\vec{q}}^{\alpha\beta,ZO}=
  [2(\gamma_1')^2\hbar^2/(M\Omega_{ZO})]\delta_{\alpha\beta} (\hbar
  v/\gamma_1)^2kp$. Here $M=A\rho_{2}$, and $\gamma_1'$ is the derivative of
  $\gamma_1$ with respect to the A1-B2 bond length. Since there is no interband
  coupling, such a mode is relatively ineffective at dissipating energy for
  $|\mu|,k_BT_e \ll \Omega_{ZO}$.}

\bibitem[{\citenamefont{Nienhaus et~al.}(1995)\citenamefont{Nienhaus, Kampen,
  and M\"onch}}]{nienhaus95}
\bibinfo{author}{\bibfnamefont{H.}~\bibnamefont{Nienhaus}},
  \bibinfo{author}{\bibfnamefont{T.~U.} \bibnamefont{Kampen}},
  \bibnamefont{and} \bibinfo{author}{\bibfnamefont{W.}~\bibnamefont{M\"onch}},
  \bibinfo{journal}{Surf. Sci.} \textbf{\bibinfo{volume}{324}},
  \bibinfo{pages}{L328 } (\bibinfo{year}{1995}).

\bibitem[{\citenamefont{Trushin and Schliemann}(2007)}]{trushin07}
\bibinfo{author}{\bibfnamefont{M.}~\bibnamefont{Trushin}} \bibnamefont{and}
  \bibinfo{author}{\bibfnamefont{J.}~\bibnamefont{Schliemann}},
  \bibinfo{journal}{Phys. Rev. Lett.} \textbf{\bibinfo{volume}{99}},
  \bibinfo{pages}{216602} (\bibinfo{year}{2007}).

\bibitem[{\citenamefont{Bolotin et~al.}(2008)\citenamefont{Bolotin, Sikes,
  Hone, Stormer, and Kim}}]{bolotin08}
\bibinfo{author}{\bibfnamefont{K.~I.} \bibnamefont{Bolotin}},
  \bibinfo{author}{\bibfnamefont{K.~J.} \bibnamefont{Sikes}},
  \bibinfo{author}{\bibfnamefont{J.}~\bibnamefont{Hone}},
  \bibinfo{author}{\bibfnamefont{H.~L.} \bibnamefont{Stormer}},
  \bibnamefont{and} \bibinfo{author}{\bibfnamefont{P.}~\bibnamefont{Kim}},
  \bibinfo{journal}{Phys. Rev. Lett.} \textbf{\bibinfo{volume}{101}},
  \bibinfo{pages}{096802} (\bibinfo{year}{2008}).

\bibitem[{\citenamefont{M\"{u}ller et~al.}(2009)\citenamefont{M\"{u}ller,
  Br\"{a}uninger, and Trauzettel}}]{muller09}
\bibinfo{author}{\bibfnamefont{M.}~\bibnamefont{M\"{u}ller}},
  \bibinfo{author}{\bibfnamefont{M.}~\bibnamefont{Br\"{a}uninger}},
  \bibnamefont{and}
  \bibinfo{author}{\bibfnamefont{B.}~\bibnamefont{Trauzettel}},
  \bibinfo{journal}{Phys. Rev. Lett.} \textbf{\bibinfo{volume}{103}},
  \bibinfo{eid}{196801} (\bibinfo{year}{2009}).

\bibitem[{\citenamefont{Adam and {Das Sarma}}(2008)}]{adam08}
\bibinfo{author}{\bibfnamefont{S.}~\bibnamefont{Adam}} \bibnamefont{and}
  \bibinfo{author}{\bibfnamefont{S.}~\bibnamefont{{Das Sarma}}},
  \bibinfo{journal}{Phys. Rev. B} \textbf{\bibinfo{volume}{77}},
  \bibinfo{pages}{115436} (\bibinfo{year}{2008}).

\bibitem[{\citenamefont{Sarma et~al.}()\citenamefont{Sarma, Hwang, and
  Rossi}}]{dassarma09}
\bibinfo{author}{\bibfnamefont{S.~D.} \bibnamefont{Sarma}},
  \bibinfo{author}{\bibfnamefont{E.~H.} \bibnamefont{Hwang}}, \bibnamefont{and}
  \bibinfo{author}{\bibfnamefont{E.}~\bibnamefont{Rossi}},
  \bibinfo{note}{arXiv:0912.0403}.

\bibitem[{\citenamefont{Morozov et~al.}(2008)\citenamefont{Morozov, Novoselov,
  Katsnelson, Schedin, Elias, Jaszczak, and Geim}}]{morozov08}
\bibinfo{author}{\bibfnamefont{S.~V.} \bibnamefont{Morozov}},
  \bibinfo{author}{\bibfnamefont{K.~S.} \bibnamefont{Novoselov}},
  \bibinfo{author}{\bibfnamefont{M.~I.} \bibnamefont{Katsnelson}},
  \bibinfo{author}{\bibfnamefont{F.}~\bibnamefont{Schedin}},
  \bibinfo{author}{\bibfnamefont{D.~C.} \bibnamefont{Elias}},
  \bibinfo{author}{\bibfnamefont{J.~A.} \bibnamefont{Jaszczak}},
  \bibnamefont{and} \bibinfo{author}{\bibfnamefont{A.~K.} \bibnamefont{Geim}},
  \bibinfo{journal}{Phys. Rev. Lett.} \textbf{\bibinfo{volume}{100}},
  \bibinfo{pages}{016602} (\bibinfo{year}{2008}).

\bibitem[{\citenamefont{Barreiro et~al.}(2009)\citenamefont{Barreiro, Lazzeri,
  Moser, Mauri, and Bachtold}}]{barreiro09}
\bibinfo{author}{\bibfnamefont{A.}~\bibnamefont{Barreiro}},
  \bibinfo{author}{\bibfnamefont{M.}~\bibnamefont{Lazzeri}},
  \bibinfo{author}{\bibfnamefont{J.}~\bibnamefont{Moser}},
  \bibinfo{author}{\bibfnamefont{F.}~\bibnamefont{Mauri}}, \bibnamefont{and}
  \bibinfo{author}{\bibfnamefont{A.}~\bibnamefont{Bachtold}},
  \bibinfo{journal}{Phys. Rev. Lett.} \textbf{\bibinfo{volume}{103}},
  \bibinfo{eid}{076601} (\bibinfo{year}{2009}).

\bibitem[{\citenamefont{Chauhan and Guo}(2009)}]{chauhan09}
\bibinfo{author}{\bibfnamefont{J.}~\bibnamefont{Chauhan}} \bibnamefont{and}
  \bibinfo{author}{\bibfnamefont{J.}~\bibnamefont{Guo}},
  \bibinfo{journal}{Appl. Phys. Lett.} \textbf{\bibinfo{volume}{95}},
  \bibinfo{eid}{023120} (\bibinfo{year}{2009}).

\bibitem[{\citenamefont{Auer et~al.}(2006)\citenamefont{Auer, Sch\"urrer, and
  Ertler}}]{auer06}
\bibinfo{author}{\bibfnamefont{C.}~\bibnamefont{Auer}},
  \bibinfo{author}{\bibfnamefont{F.}~\bibnamefont{Sch\"urrer}},
  \bibnamefont{and} \bibinfo{author}{\bibfnamefont{C.}~\bibnamefont{Ertler}},
  \bibinfo{journal}{Phys. Rev. B} \textbf{\bibinfo{volume}{74}},
  \bibinfo{pages}{165409} (\bibinfo{year}{2006}).

\bibitem[{\citenamefont{Lazzeri and Mauri}(2006)}]{lazzeri06}
\bibinfo{author}{\bibfnamefont{M.}~\bibnamefont{Lazzeri}} \bibnamefont{and}
  \bibinfo{author}{\bibfnamefont{F.}~\bibnamefont{Mauri}},
  \bibinfo{journal}{Phys. Rev. B} \textbf{\bibinfo{volume}{73}},
  \bibinfo{pages}{165419} (\bibinfo{year}{2006}).

\bibitem[{\citenamefont{Song et~al.}(2008)\citenamefont{Song, Wang, Dukovic,
  Zheng, Semke, Brus, and Heinz}}]{song08}
\bibinfo{author}{\bibfnamefont{D.}~\bibnamefont{Song}},
  \bibinfo{author}{\bibfnamefont{F.}~\bibnamefont{Wang}},
  \bibinfo{author}{\bibfnamefont{G.}~\bibnamefont{Dukovic}},
  \bibinfo{author}{\bibfnamefont{M.}~\bibnamefont{Zheng}},
  \bibinfo{author}{\bibfnamefont{E.~D.} \bibnamefont{Semke}},
  \bibinfo{author}{\bibfnamefont{L.~E.} \bibnamefont{Brus}}, \bibnamefont{and}
  \bibinfo{author}{\bibfnamefont{T.~F.} \bibnamefont{Heinz}},
  \bibinfo{journal}{Phys. Rev. Lett.} \textbf{\bibinfo{volume}{100}},
  \bibinfo{pages}{225503} (\bibinfo{year}{2008}).

\end{thebibliography}

\end{document}